\documentclass[12pt,a4paper,english,superscriptaddress,aps,nofootinbib]{revtex4}
\usepackage[utf8]{inputenc}
\usepackage[T1]{fontenc}
\usepackage{amsmath,amssymb,graphicx}
\makeatletter
\usepackage{babel}
\usepackage[active]{srcltx}
\usepackage{graphicx,color}
\usepackage{changebar}
\usepackage{hyperref}
\hypersetup{
	colorlinks=true,
	urlcolor= blue,
	citecolor=blue,
	linkcolor= blue,
	bookmarks=true,
	bookmarksopen=false,}

\usepackage[T1]{fontenc}
\usepackage{ae}
\usepackage{amsmath}
\usepackage{braket}
\usepackage{dsfont}
\usepackage{mathtools}
\usepackage{slashed}
\usepackage{empheq}
\usepackage{tikz}
\usepackage{subfigure}
\usepackage{multirow}
\begin{document}	
 
 \title{Asymmetric domain walls in modified $\phi^4$ theory: Excitation spectra, scattering, and decay of bions}

\author{F. C. E. Lima}
\email[]{E-mail: cleiton.estevao@fisica.ufc.br}
\affiliation{Departamento de F\'{i}sica, Universidade Federal do Cear\'{a} (UFC), Campus do Pici, Fortaleza - CE, 60455-760, Brazil.}

\begin{abstract}
\vspace{0.5cm}

\noindent \textbf{Abstract:} We consider a two-dimensional Lorentz-invariant field model with a $\phi^4$ potential modified by a term that introduces asymmetries at the manifold space. In this framework, the model recovers its original symmetry only when $p=0$. The asymmetry introduced in the potential suggests that, even when one of the minima diverges asymptotically, kink/antikink-like configurations emerge in the theory, shifting the critical point of the energy density away from the center of the kink-like solutions. Hence, we note that the model supports asymmetrical kink/antikink-like topological solutions. Furthermore, an analysis of the excitation spectrum of these solutions revealed the absence of vibrational modes. Finally, we examine the dynamical solutions for different values of initial velocity by allowing us to verify the effects of asymmetries on the collision properties.
\end{abstract}

\maketitle

\thispagestyle{empty}

\newpage 

\section{Introduction}

Topological defects are configurations originating from spontaneous symmetry breaking, characterized by a structure associated with a non-trivial topological space\footnote{Nontrivial topological spaces exhibit remarkable properties such as connectivity and the presence of ``holes'' - for instance, ($S^1$) in two-dimensional and ($S^2$) in three-dimensional theories. We classify a space as connected or non-simply connected. Formally, the nontrivial nature of such spaces is often associated with nonzero homotopy or homology groups, which highlight the existence of specific topological structures. Notable examples of these structures include domain walls (or kinks), which display invariance characteristics of nontrivial topological spaces, such as circles ($S^1$) or torus ($S^2$). These structures persist even under continuous deformations, reflecting the robustness of their topology  \cite{Vachaspati,Rajaraman,Manton}.} \cite{Vachaspati,Rajaraman,Manton}. In this context, the manifold topology plays a fundamental role in the characterization of the classes of defects. Among the most extensively studied topological defects are domain walls, cosmic strings, and monopoles \cite{Vilenkin}. In this study, we will focus on the analysis of domain walls. It is essential to highlight that the domain walls (or kinks) are the most elementary example of topological defects. These objects arise, in particular, in the context of classical field theories that aim to describe the essential properties for the existence of spin-1/2 particles \cite{Finkelstein}. Besides, kinks play a significant role in modern physics, as they appear in various phenomena \cite{Romanczukiewicz}, such as the formation of magnetic domain walls \cite{Glathe,Buijnsters} and structural defects in crystals \cite{Ishizaka,Gou}. Surprisingly, these configurations also play a crucial role in the condensation process of DNA into the nucleosome \cite{Olson}. Despite numerous applications and considerable progress in the study of this topological defect over the past decades \cite{Malomed,Dorey,Christov}, several open questions remain. One of the key challenges is developing more robust techniques to describe the deformation of these structures and gain a deeper understanding of their dynamics.

Among the various models capable of giving rise to kink configurations, $\phi^4$ models stand out prominently in the literature. These theories hold significant importance once they describe phase transitions in several contexts, such as the Ginzburg-Landau theory \cite{Ginzburg,Landau}. Furthermore, the $\phi^4$ model has numerous additional applications, see Refs. \cite{Kevrekidis,Saadatmand}. A particularly notable example is the application of kinks in describing structural deformations in graphene \cite{Yamaletdinov1,Yamaletdinov2}. In this framework, one employs the polynomial theories, i.e., the $\phi^4$-like model. Indeed, one can note that the $\phi^4$ model is widely used to investigate phase transitions and their properties \cite{AKhare,Pavlov,Mroz}. Recently, models with non-polynomial potentials have shown effectiveness in the first-order phase transition process between lamellar and inverse hexagonal phases in specific lipid bilayers \cite{Mendanha}. Encouraged by these breakthroughs, we adopt a $\phi^4$ theory modified with an additional term responsible  the emergence of asymmetry phenomenon. This modification introduces greater rigidity near the vacuum, making such models particularly attractive due to their ability to produce more adaptable kink configurations. This adaptability, in turn, can significantly influence the structural resilience of ferroelectric materials \cite{Bazeia}. To explore these potentially more flexible configurations, we will investigate a theory in which asymmetry plays a central role. This phenomenon is of particular interest as it contributes to the localization (or non-localization) of the theory’s vacua, leading to structural asymmetries that affect the dynamics of the solutions. Therefore, our primary goal is to understand the effects of the asymmetry and its implications for the structure.

In this work, we introduce the scalar field theory and modify the $\phi^4$ potential by implementing an asymmetry into the vacuum states $\phi = -\nu$ (on the left) and $\phi = +\nu$ (on the right). Additionally, one analyzes the static solutions by varying the asymmetry. Furthermore, we investigate the excitation spectrum and zero mode. Finally, we are devoted to studying the dynamic case and announcing our findings.
    
\section{The standard scalar field theory}\label{secII}

We shall commence our project by considering the two-dimensional action of a scalar field theory $\phi(x,t)$ in flat two-dimensional spacetime, viz.,
\begin{align}
    \label{Eq1}
    S=\int\,d^2x\,\left[\frac{1}{2}\partial_\mu\phi\,\partial^\mu\phi-V(\phi)\right].
\end{align}
In this framework, $V(\phi)$ represents the potential, and the term $\frac{1}{2}\partial_\mu\phi\partial^\mu\phi$ corresponds to the kinetic contribution of the scalar theory. Furthermore, $x$ is the spatial coordinate, and $t$ is the temporal variable. In this study, we adopt the metric signature $\eta_{\mu\nu}=(+,-)$.

To investigate the equation of motion, let us apply the principle of least action ($\delta S = 0$), which leads us to
\begin{align}
    \label{Eq2}
    \partial^{\mu}\partial_\mu\phi+V_\phi=0,
\end{align}
where $V_\phi=\frac{\partial V}{\partial\phi}$. Naturally, we can reformulate the expression \eqref{Eq2} as
\begin{align}\label{Eq3}
    \Ddot{\phi}-\phi''+V_\phi=0.
\end{align}
Here, the dot and prime notations are the derivatives concerning the time and position, respectively.

Adopting the static case, i.e., $\dot{\phi} = 0$\footnote{The study of the static case is advantageous once it is possible to derive the bsolution of $\phi(x,t)$ by applying a Lorentz boost on the static field solution. For further details, see Refs. \cite{Vachaspati,Rajaraman}.}, and multiplying all terms from Eq. \eqref{Eq3} by $\phi'$, one arrives at
\begin{align}\label{Eq4}
    \frac{d\phi}{dx}\mp\sqrt{2V(\phi)}=0.
\end{align}
Eq. \eqref{Eq4} is the BPS\footnote{The term BPS refers to the physicists Bogomol'nyi \cite{Bogomolnyi}, Prasad, and Sommerfield \cite{PS}, who are responsible for formulating the approach to obtain the field classical solutions. Briefly, the BPS approach describes stable solutions or low-energy configurations that preserve the theory's symmetry by reducing the order of the equation of motion.} equation for the static scalar field.

In the static regime, the scalar field energy is
\begin{align}
    \label{Eq5}
    \textrm{E}=\int\, dx\, \left[\frac{1}{2}\left(\frac{d\phi}{dx}\right)^2+V\right].
\end{align}

By substituting Eq. \eqref{Eq4} into Eq. \eqref{Eq5}, it follows that
\begin{align}\label{Eq6}
    \textrm{E}_{\textrm{BPS}}=\int\, 2V[\phi(x)]\,\, dx,
\end{align}
which allows us to announce that the BPS energy density of the scalar field configurations is
\begin{align}\label{Eq7}
    \rho_{\textrm{BPS}}=2V[\phi(x)].
\end{align}

To proceed with the study, one must adopt a potential $V(\phi)$ for the theory announced in Eq. \eqref{Eq1}. We will assume the modified $\phi^4$ theory is a low-energy description for specific microscopic models. Such descriptions are justified because the imperfections at the microscopic level are associated with variations in the potential \cite{Lizunova,Abdullaev,Bilas,Zhou}. A relevant example is waves propagating through a solid medium, where impurities or defects arise in the crystalline structure and induce scattering of incident waves \cite{Lizunova}. Furthermore, let us modify our theory for a non-polynomial theory, i.e., with hyperbolic contributions to ensure the formation of kink-like configurations. We will perform this because these configurations are relevant in descriptions of ferroelectric and ferromagnetic materials. Besides, the hyperbolic functions arise because models described by polynomial and Sine-Gordon-like interactions exhibit substantial fragility owing to the rigidity of certain ferroelectric materials \cite{Bazeia}. For further details, see Refs. \cite{Kofane1,Kofane2}. Other models of hyperbolic applications are similarly integrable extensions of the Calogero model \cite{Calogero}, generating multi-soliton solutions in confined potentials \cite{Gon}. Potentials with hyperbolic contributions also arise in $\mathcal{N}=2$ and $\mathcal{N}=4$ supersymmetric theories \cite{Fedoruk}, hairy black holes \cite{Wen}, and quintessential inflation \cite{Agarwal}. Thus, we adopt the potential
\begin{align}
    \label{Eq9}
    V(\phi)=\frac{\lambda}{2}(\nu^2-\phi^2)^2\,\textrm{e}^{2p\,\phi}=\frac{\lambda}{2}(\nu^2-\phi^2)^2[\sinh(2p\phi)+\cosh(2p\phi)].
\end{align}
where $p\in \mathds{R}$. This interaction describes the modified $\phi^4$ potential by an exponential global impurity $\textrm{e}^{2 p \phi}$. Note that when $p=0$, we recover the na\"{i}ve $\phi^4$ theory\footnote{Furthermore, one notes that for $p$ (asymmetry) very small, we recovered the standard $\phi^4$ theory.}. Meanwhile, if $p \neq 0$, it will produce an asymmetry in the neighborhood of one of the vacua \cite{GaniJHEP,FCELima}. In this sense, when $p>0$ will produce an asymmetry (or vacua asymmetric) around the vacuum $-\nu$ (left), and $p<0$ produces decay around the vacuum $+\nu$ (right).

\begin{figure}[!ht]
  \centering
  \includegraphics[height=7cm,width=8cm]{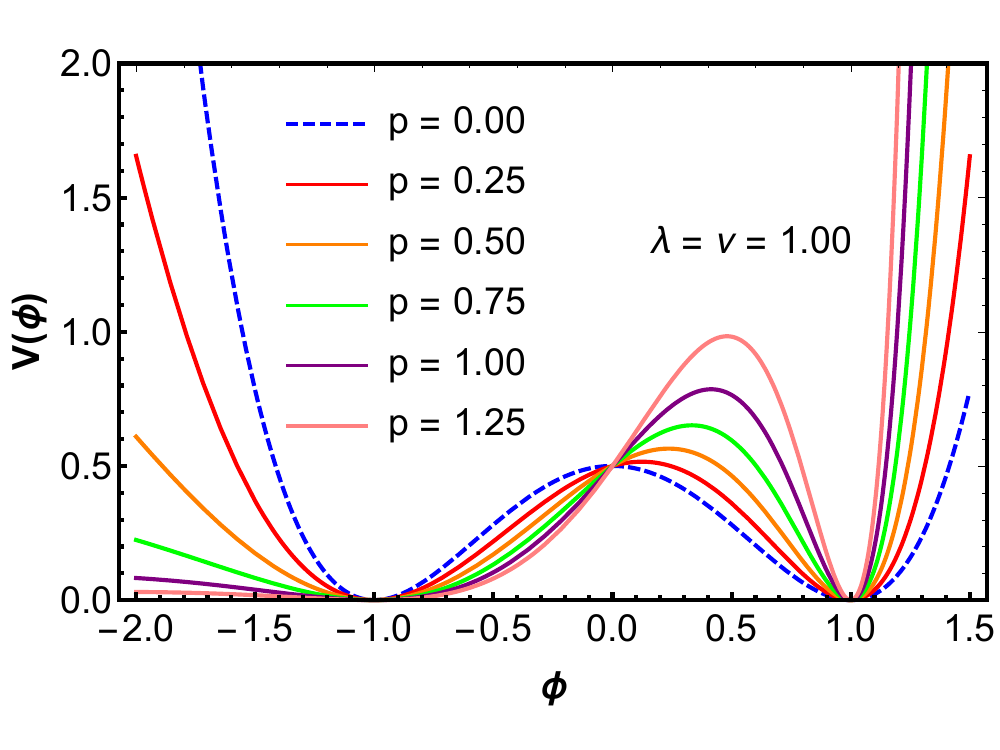}\vspace{-0.3cm}
  \caption{Plot of the potential $V(\phi)$ vs. $\phi$. We build this plot adopting $\lambda=\nu=1.00$.}  \label{fig1}
\end{figure}

Based on the potential described by Eq. \eqref{Eq9}, let us study the system by considering the asymmetry around $\phi = -\nu$ (asymmetry on the left). We present the plot corresponding to the potential \eqref{Eq9} in Fig. \ref{fig1}.

In this framework, the equation of motion is 
\begin{align}
    \label{Eq10}
    \phi'(x)\mp\sqrt{\frac{\lambda}{2}}[\nu^2-\phi(x)^2]\,\textrm{e}^{p\,\phi(x)}=0.
\end{align}
Note that, when $p=0$, the solutions of Eq. \eqref{Eq10} can be found analytically, namely,
\begin{align}\label{Eq11}
    \phi(x)=\pm \nu\tanh\left(\sqrt{\frac{\lambda}{2}}\nu x\right).
\end{align}
These solutions are called true kinks (positive sign) and antikinks (negative sign).

%-------------------------------------------------------

\subsection{The case with the asymmetry on the left ($p>0$)}\label{secIIa}

\begin{figure}[!ht]
  \centering
  \subfigure[Asymmetric kink-like solutions.]{\includegraphics[height=7cm,width=8cm]{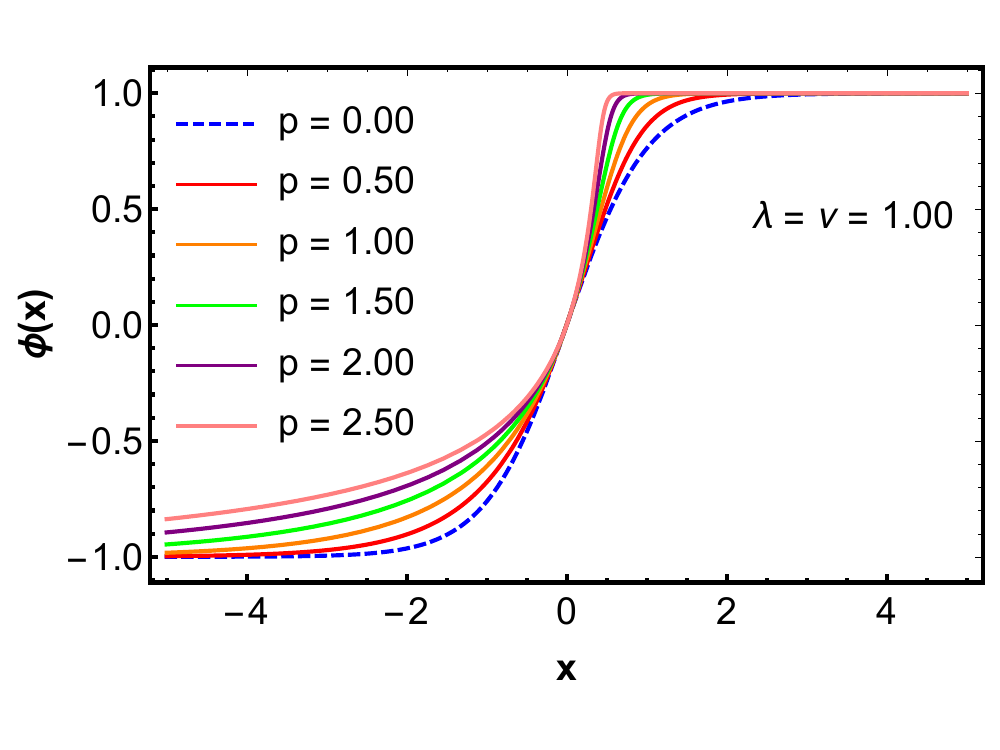}}\hfill
  \subfigure[Asymmetric antikink-like solutions.]{\includegraphics[height=7cm,width=8cm]{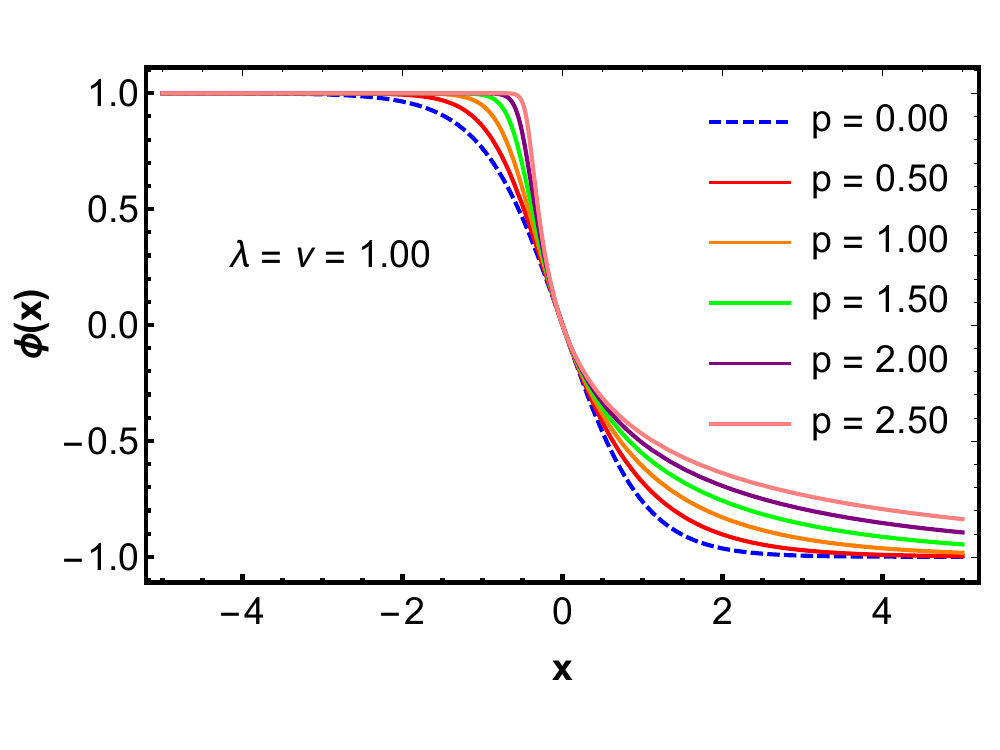}}\hfill
  \caption{Numerical solutions of the scalar field $\phi$ vs. position ($x$) with asymmetry on the left. We build this plot adopting $\lambda=\nu=1$.}  \label{fig2}
\end{figure}

For the case where $p> 0$, it is necessary to conduct a numerical study of Eq. \eqref{Eq10}. Using the numerical interpolation method, we study the numerical solution of Eq. \eqref{Eq10}. One exposes the numerical solutions in Figs. \ref{fig2}(a) and \ref{fig2}(b). From the analysis of the numerical solutions [Figs. \ref{fig2}(a) and \ref{fig2}(b)], we infer that the solutions represent kink-like configurations connecting the vacuum belonging to the topological sectors $(-\nu, \nu)$. Thus, the BPS solution of Eq. \eqref{Eq10} interpolates between the vacuum $\phi_{-\infty} = -\nu$ and $\phi_\infty = \nu$. Furthermore, there are also antikink-like solutions corresponding to the same vacua but interpolating between $\phi_{-\infty} = \nu$ and $\phi_\infty = -\nu$. These configurations belong to the same topological sector $(-\nu, \nu)$. In summary, the distinction between kink- and antikink-like configurations is only a convention. Besides, one highlights a characteristic of this class of solutions, which is modified with the asymmetry, i.e., while kink/antikink solutions ($p = 0$) are symmetric around $x = 0$, for the case $p>0$, this symmetry is broken. Indeed, one can note that the asymmetry on the left shortens (or deforms) the top of the solutions, resulting in configurations with a more compact-like profile at their top ($0<\phi\leq\nu$). We call these configurations of asymmetrical kink/antikink-like solutions.

It is interesting to highlight that structures (Fig. \ref{fig2}) resembling arise in higher-order theory in a two-dimensional flat spacetime. For instance, similar configurations appear in some topological sectors of the $\phi^8$ theory \cite{GaniJHEP}. Furthermore, one notes the emergence of these asymmetric structures in the description of the matter sector in extra-dimensional theories (see Ref. \cite{FCELima}). In both cases, this asymmetry manifests in the neighborhood of one of the vacua. 

\begin{figure}[!ht]
  \centering
  \includegraphics[height=7cm,width=8cm]{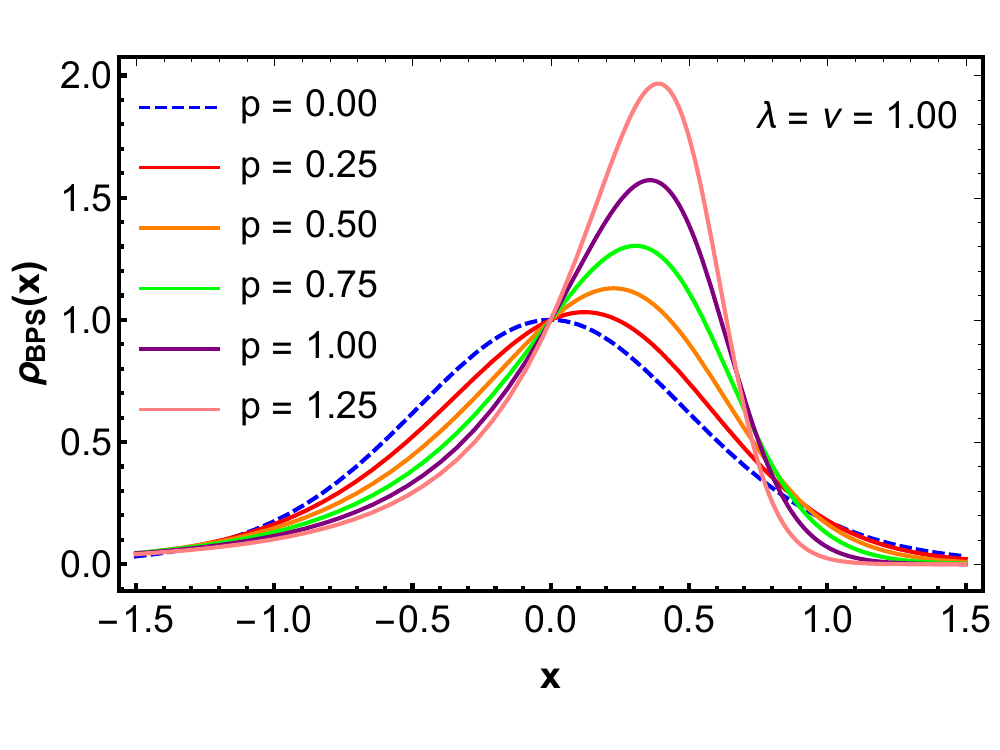}\vspace{-0.3cm}
  \caption{Energy density $\rho_{\textrm{BPS}}(x)$ vs. $x$ when $\lambda=\nu=1$.}  \label{fig3}
\end{figure}

Now, allow us to examine the BPS energy density. We begin our analysis by reformulating the Eq. \eqref{Eq10} as 
\begin{align}\label{Eq12}
    \rho_{\textrm{BPS}}= \lambda(\nu^2-\phi^2)^2\textrm{e}^{2p\,\phi},
\end{align}
and using the numerical solutions presented in Figs. \ref{fig2}(a) and \ref{fig2}(b), we perform a numerical investigation of the BPS energy density associated with the solutions of the Eq. \eqref{Eq10}. The BPS energy density as a function of the position $x$ is presented numerically in Fig. \ref{fig3}.
Upon analyzing the behavior of the energy density (Fig. \ref{fig3}), one notes that the asymmetry on the left produced a compactification on the scalar field and, consequently, the emergence from critical energy point more intense. Furthermore, this result resonates in the energy density by shifting the energy from $x=0$ to the right with increasing intensity as the parameter $p$ grows. This behavior of the energy confirms the asymmetric nature of the structures.

%--------------------------------------------------------------------------------

\subsection{The case with the asymmetry on the right ($p<0$)}\label{secIIb}

\begin{figure}[!ht]
  \centering
  \includegraphics[height=7cm,width=8cm]{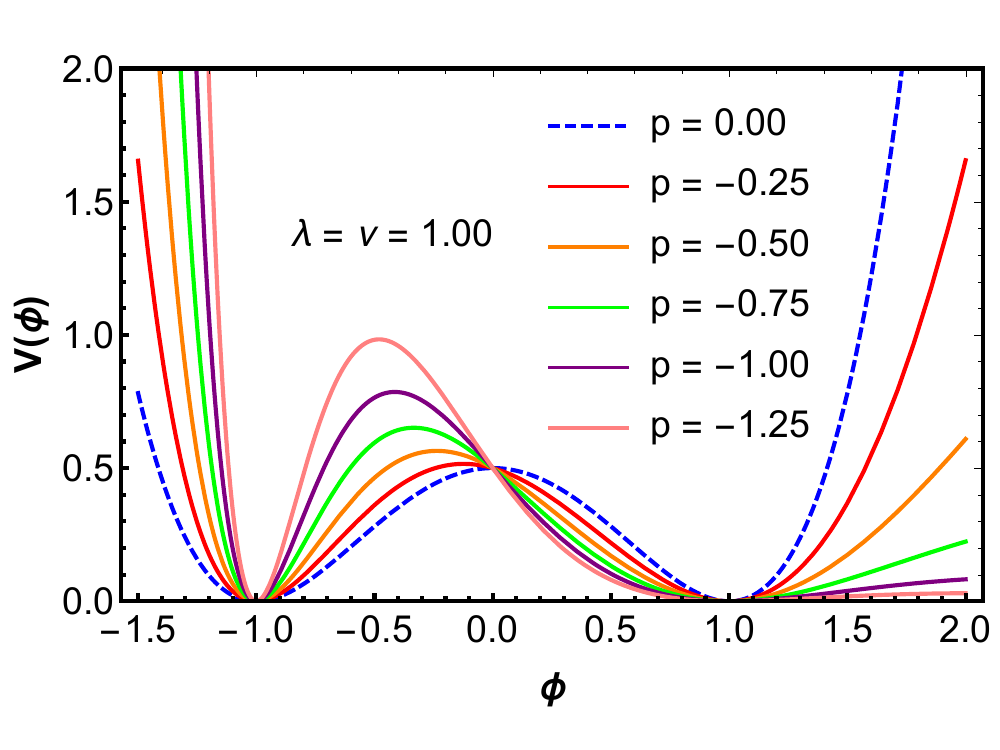}\vspace{-0.3cm}
  \caption{Potential $V(\phi)$ vs. $\phi$. We build this plot adopting $\lambda=\nu=1$. In this case, one exposes several curves for different $p$ values.}  \label{fig4}
\end{figure}

Once we investigated the case with the asymmetry on the left ($p>0$), allow us to examine the case with the asymmetry on the right ($p<0$). In this case, the potential vs. matter field assumes the form announced in Fig. \ref{fig4}. Naturally, we note the asymmetry on the potential near $\phi = \nu$, which breaks the symmetry at $\phi=\nu$.

\begin{figure}[!ht]
  \centering
  \subfigure[Asymmetric kink-like solutions.]{\includegraphics[height=7cm,width=8cm]{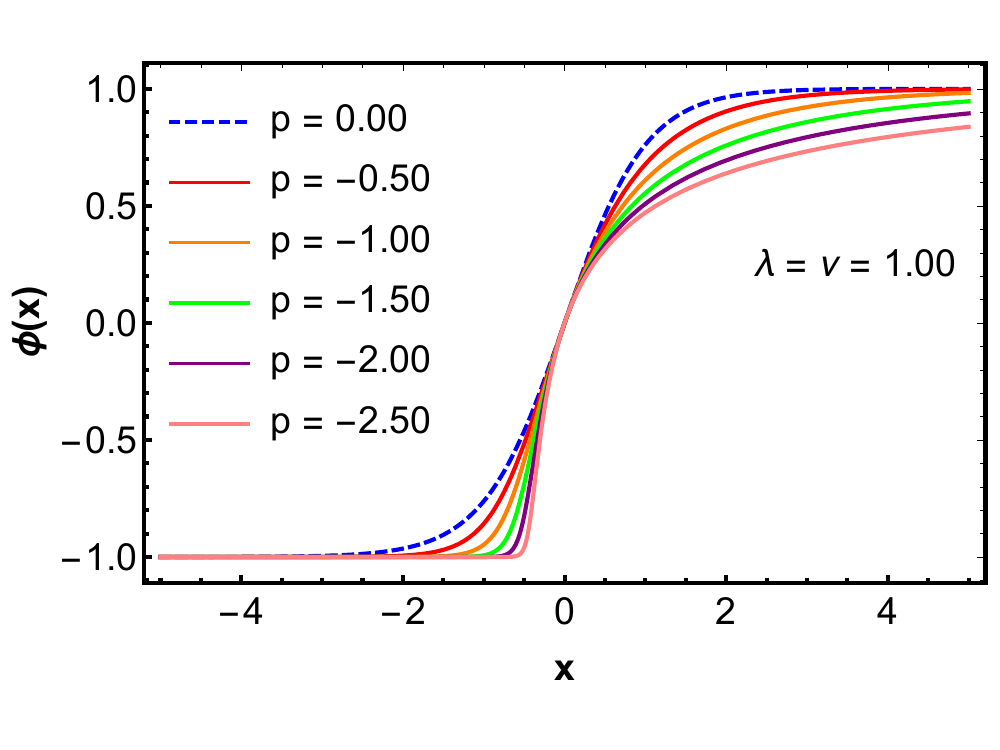}}\hfill
  \subfigure[Asymmetric antikink-like solutions.]{\includegraphics[height=7cm,width=8cm]{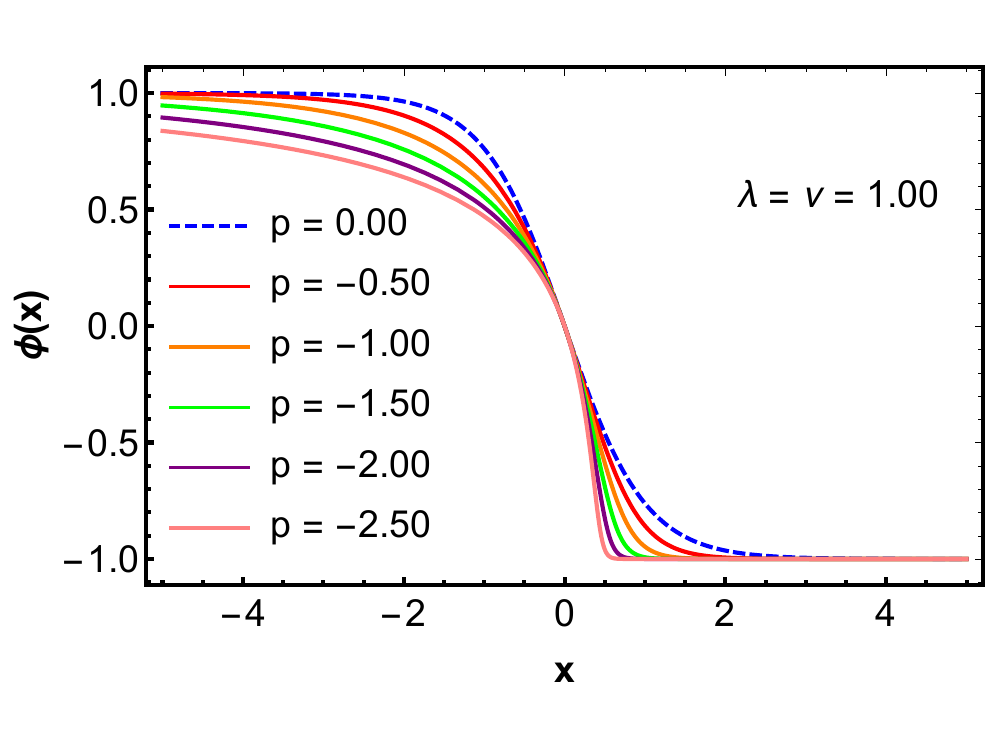}}\hfill
  \caption{Numerical solutions of the scalar field $\phi$ vs. position ($x$) with asymmetry on the right. We build this plot adopting $\lambda=\nu=1$.}  \label{fig5}
\end{figure}

Considering the case with asymmetry on the right, we solve Eq. \eqref{Eq10} by adopting $p<0$. Again, we need to use a numerical approach to obtain the solutions of Eq. \eqref{Eq10}. Thus, we will apply the numerical interpolation method to reach our purpose. The solutions for this case are presented in Fig. \ref{fig5}.

Examining the solution for the case $p<0$, we noted that the solutions are again kink/antikink-like configurations connecting the vacua belonging to the topological sector $(-\nu,\nu)$. Therefore, these solutions interpolate between the vacua $\phi_{-\infty}=-\nu$ and $\phi_\infty=\nu$, with a kink-like profile resembling the semi-compact-like.\footnote{We call semi-compact-like configurations the structures that shrink only on one side concerning the initial position of the kink/antikink-like solution, forming a corner-like profile.}. Meanwhile, there will also be antikink-like configurations (resembling semi-compact) corresponding to the same vacuum. However, these solutions will interpolating between $\phi_{-\infty}=\nu$ and $\phi_{\infty}=-\nu$. Furthermore, it is noteworthy that the solutions obtained for the asymmetry on the right modify the true kink/antikink solutions ($p=0$), rendering them asymmetric around $x=0$. In summary, this asymmetry arises due to the asymmetry around the potential minimum at $\phi=\nu$ (asymmetry on the right). It is essential to highlight that, unlike the case with asymmetry on the left, the solutions exhibit profiles more compacted at the lower part of the structure (or solutions), i.e., in $-1\leq\phi<0$.

\begin{figure}[!ht]
  \centering
  \includegraphics[height=7cm,width=8cm]{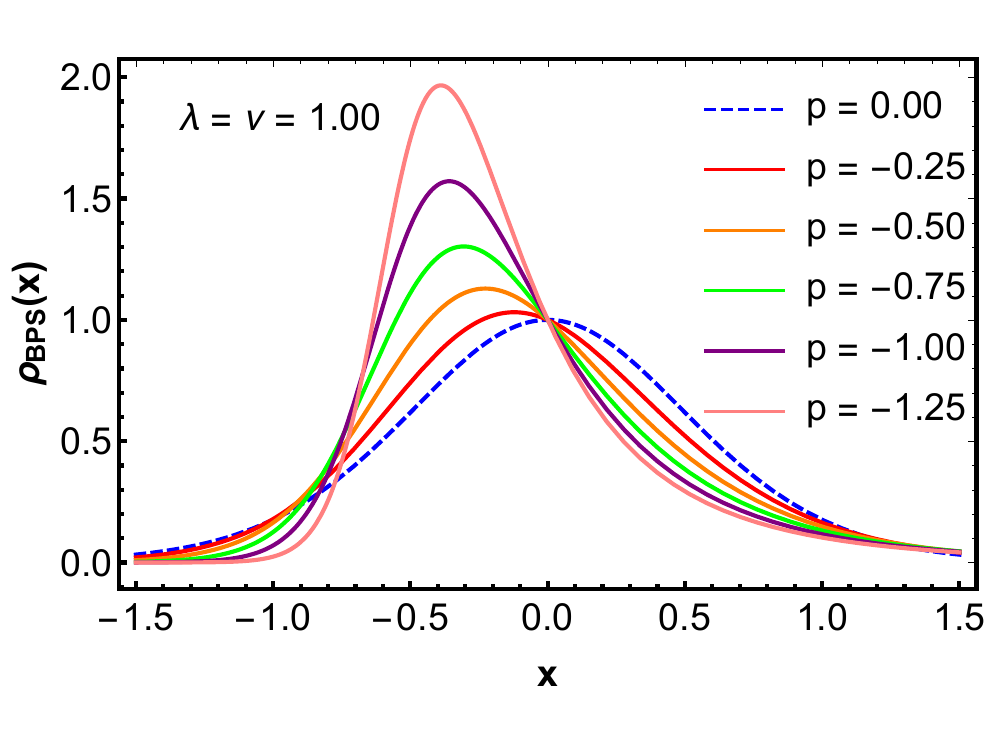}\vspace{-0.4cm}
  \caption{Energy density $\rho_{\textrm{BPS}}(x)$ vs. $x$ when $\lambda=\nu=1$.}  \label{fig6}
\end{figure}

To conclude the discussion of the system's solutions, we will analyze the BPS energy density \eqref{Eq12} related to the solutions depicted in Figs. \ref{fig5}(a) and \ref{fig5}(b). To carry out this analysis, we will substitute the numerical solutions [Figs. \ref{fig5}(a) and \ref{fig5}(b)] into the expression \eqref{Eq12}. Thus, we obtain the BPS energy density announced in Fig. \ref{fig6}.

The results presented in Fig. \ref{fig6} allow us to note that when the asymmetry on the right occurs, the scalar field feels a compactification at the base of the structure (i.e., $0\leq\phi<\infty$). Naturally, this deformation shifts the energy density at $x=0$ to the left, with the intensity of the energy density increasing as the parameter $\vert p\vert$ grows. That happens because increasing the parameter $p$ leads to a compactification of the scalar field (at the base of the structure). This result helps us to confirm the existence of asymmetric kink/antikink-like structures.

\section{Excitation spectrum of the asymmetric domain walls}\label{secIII}

In this section, we will examine the excitation spectrum of the scalar field solutions\footnote{Here, $\phi_k(x)$ describes the kink/antikink-like solutions.}. To achieve our purpose, allow us to consider a small perturbation $\delta\phi(x,t)$, viz.,
\begin{align}
    \label{Eq16}
    \phi(x,t)=\phi_k(x)+\delta \phi(x,t),
\end{align}
in the previously presented static solutions.

Applying the perturbation \eqref{Eq16} into \eqref{Eq3}, one obtains
\begin{align}
    \label{Eq17}
    -\eta''(x)+V_{\textrm{eff}}\,\eta(x)=\omega^2\eta(x),
\end{align}
where $\omega$ is the eigenvalue of the excitation spectrum, and $\eta(x)$ and $V_{\textrm{eff}}$ be the corresponding eigenfunction and effective potential (or stability potential), respectively. Additionally, the effective potential and the translational mode ($\omega=0$) are, respectively,
\begin{align}
    \label{Eq18}
    V_{\textrm{eff}}=\frac{\partial^2V}{\partial\phi^2}\bigg\vert_{\phi=\phi_k}
\end{align}
and
\begin{align}
    \label{Eq19}
    \eta_{0}=\frac{d\phi_k(x)}{dx}.
\end{align}

\subsection{The case $p\geq0$} \label{secIIIa}

\begin{figure}[!ht]
  \centering
  \subfigure[Effetive potential $V_{\textrm{eff}}$ vs. $x$.]{\includegraphics[height=7cm,width=8cm]{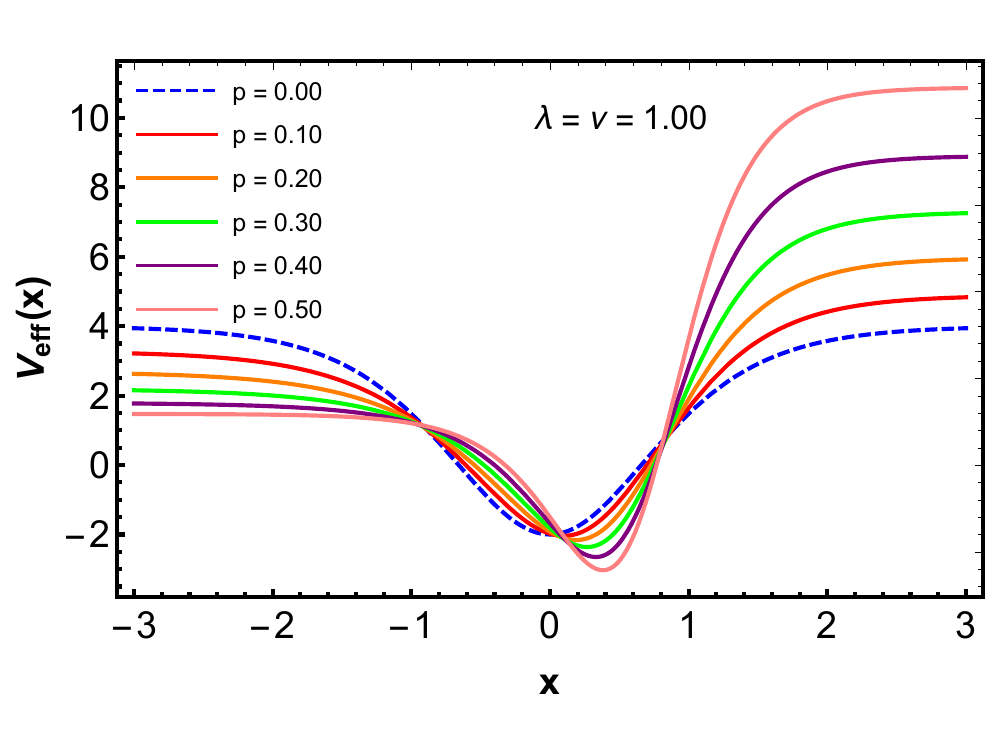}}\hfill
  \subfigure[Translational modes vs. $x$]{\includegraphics[height=7cm,width=8cm]{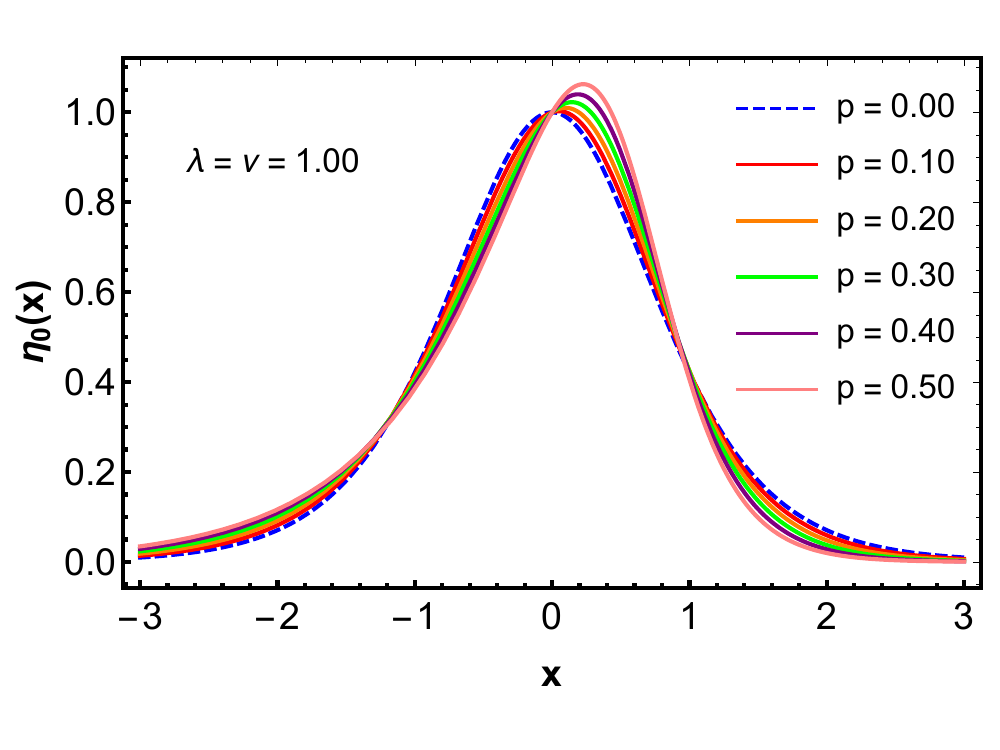}}\hfill
  \caption{Numerical solutions of the Schr\"{o}dinger-like equation \eqref{Eq17} with effective potential \eqref{Eq18}.}  \label{fig7}
\end{figure}

From this point onward, we shall examine the excitation spectrum of the $\phi^4$ model with an exponential global impurity ($\textrm{e}^{2p\phi}$), which induces the asymmetry on the left. To investigate the excitation spectrum, one replaces the numerical solutions presented in Fig. \ref{fig2} into Eq. \eqref{Eq17}. Thus, we show the numerical results of the effective potential \eqref{Eq18} and the translational mode \eqref{Eq19}, respectively, in Figs. \ref{fig7}(a) and (b).

Upon analyzing the results shown in Figs. \ref{fig7}(a) and \ref{fig7}(b), we note that the effective potential is asymmetric, which promotes the emergence only of asymmetric translational modes ($\omega = 0$).

\subsection{The case $p\leq 0$}\label{secIIIb}

\begin{figure}[!ht]
  \centering
  \subfigure[Effetive potential $V_{\textrm{eff}}$ vs. $x$.]{\includegraphics[height=7cm,width=8cm]{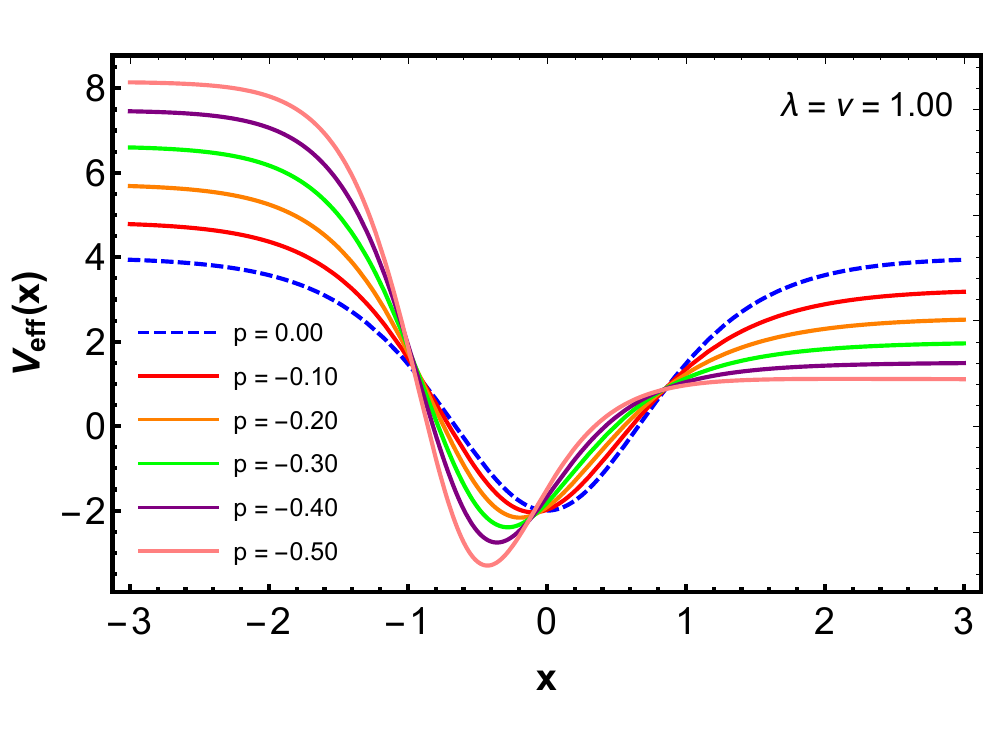}}\hfill
  \subfigure[Translational modes vs. $x$.]{\includegraphics[height=7cm,width=8cm]{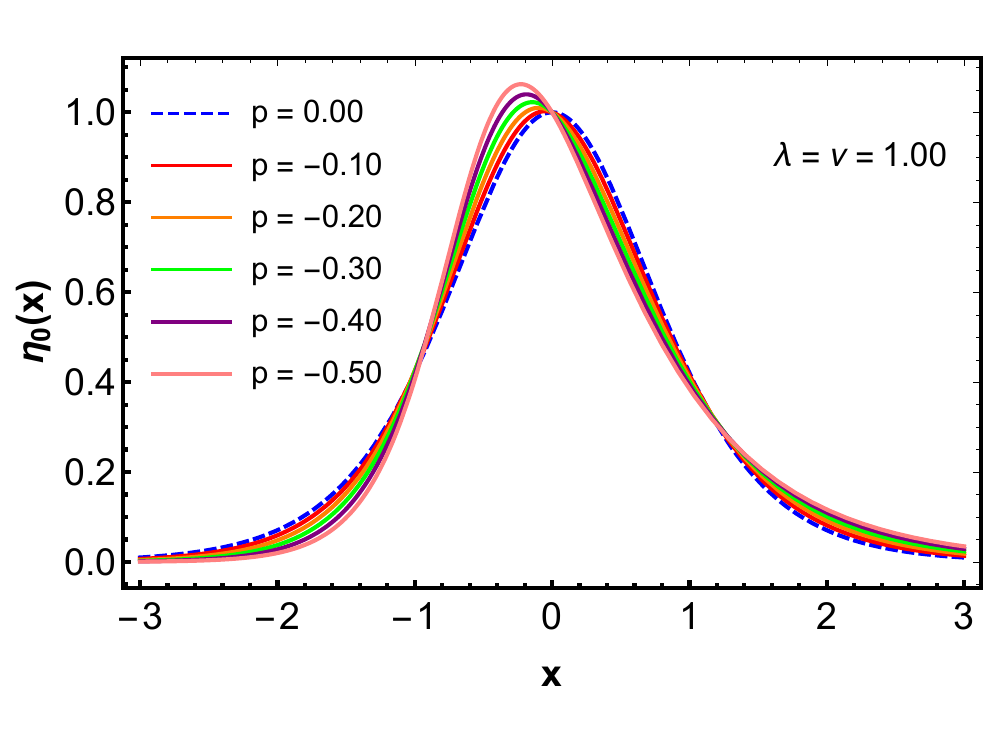}}\hfill
  \caption{Numerical solutions of the Schr\"{o}dinger's equation \eqref{Eq17} with effective potential \eqref{Eq18}.}  \label{fig8}
\end{figure}

Using a similar approach exposed in the case $p\geq 0$, we obtain the effective potential \eqref{Eq18} and the translational mode \eqref{Eq19} described in Figs. \ref{fig8}(a) and \ref{fig8}(b), respectively.

Unlike the case with asymmetry on the left, when decay occurs to the right, an asymmetric effective potential of stability arises, resulting in the shift of the potential minimum to the left, in contrast to the previous case. This behavior again produces asymmetric translational modes ($\omega=0$). One notes that the model only induces translational modes, leading to the hypothesis that collisions between configurations resembling semi-compact kink-antikink structures may transfer kinetic energy to the vibrational mode of the composite kink-antikink system. Further details on the scattering of these structures, with asymmetry on the right and left, are discussed in section \ref{secIV}.

\section{Scattering of structures}\label{secIV}

In this moment, we seek to investigate the scattering of kink-antikink solutions discussed in the previous section. To accomplish our purpose, allow us to conduct a numerical analysis of collisions involving kink/antikink-like configurations, varying the asymmetry parameter, i.e., the parameter $p$. To study the scattering process, one employs the equation of motion \eqref{Eq3} and adopts the condition\footnote{For further details on the collision of structures, see Refs. \cite{Rev1,Rev2,Rev3,Rev4}.}$^{,}$\footnote{In this conjecture, we consider the scattering of kink-antikink configurations evolving in space-time by collective coordinates. However, it is essential to highlight that the collective coordinates of kink-antikink produce an asymmetric central barrier. Meanwhile, the scattering of antikink-kink is an asymmetric central well, which leads to similar results but with the scattering of structures to the right. Thus, to avoid similar results, we will focus our study on the case of kink-antikink structures.}
\begin{align}
    \label{Eq20}
    \phi(x,t)=\phi_{K}[\gamma(x+x_0-v_{\textrm{in}}t)]+\phi_{\tilde{K}}[\gamma(x-x_0+v_{\textrm{in}}t)]-1.
\end{align}

In this framework, the indices $K$ and $\tilde{K}$ describe the kink- and antikink-like solutions, respectively. The parameter $x_0$ represents the initial position of the configurations. For convenience, we adopt the initial position being $x_0 = 5$. Besides, $v_{\textrm{in}}$ denotes the initial velocity of the structures, and $\gamma$ is the Lorentz factor.

\subsection{The case $p\geq 0$}\label{secIVa}

The evolution of the initial kink/antikink-like configurations $(-1,1,-1)$ is obtained numerically by adopting the equation of motion \eqref{Eq3}. In this framework, we note that the evolution of the configurations exhibits are independent of the asymmetry parameter. The case with asymmetry on the left induces a shift in that direction during the evolution of the structures as time progresses.

\begin{figure}[!ht]
  \centering
  \subfigure[The case $p=0.0$ and $v_{\text{in}}=0.049.$]{\includegraphics[height=4.33cm,width=5.33cm]{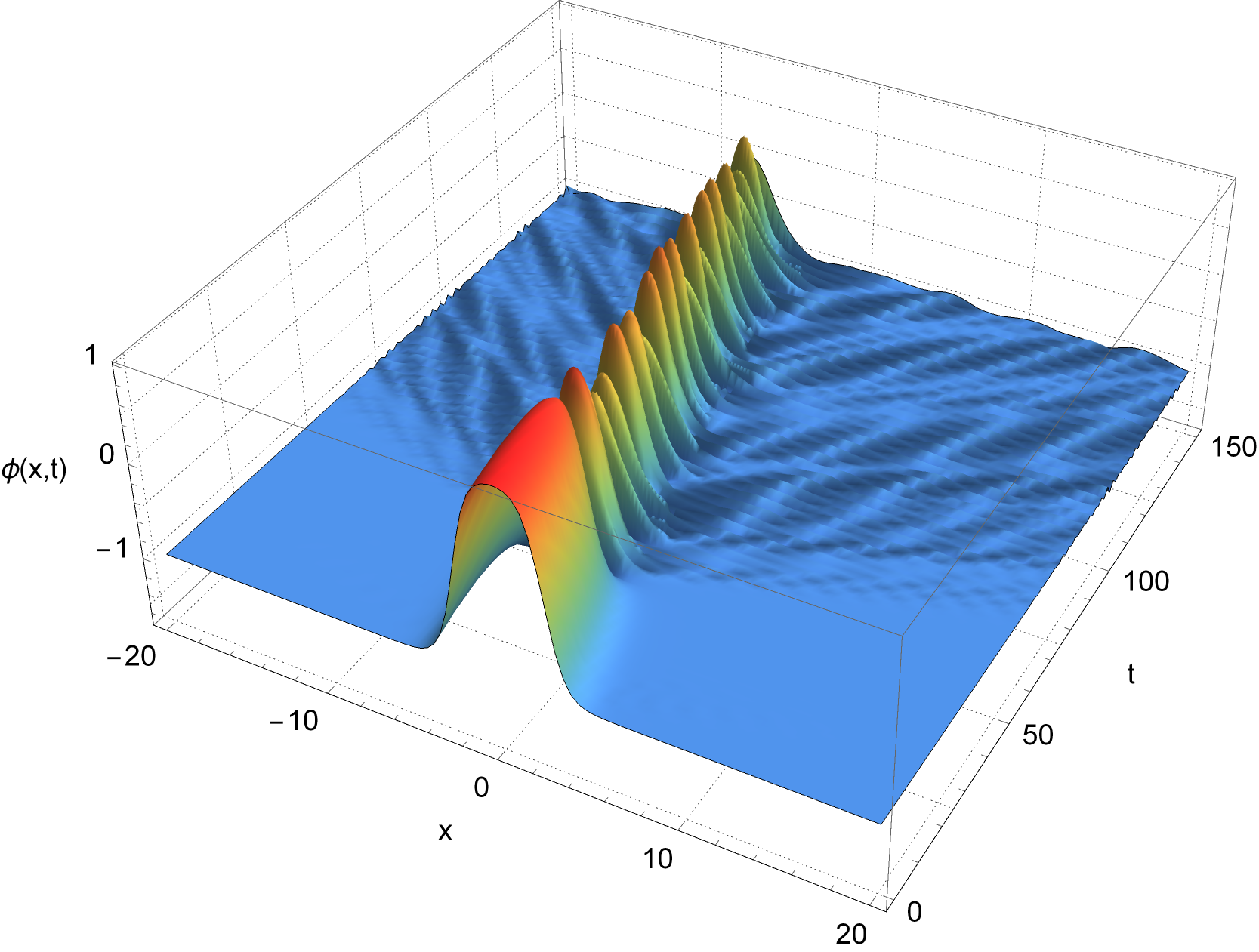}}\hfill
  \subfigure[The case $p=0.0$ and $v_{\text{in}}=0.098.$]{\includegraphics[height=4.33cm,width=5.33cm]{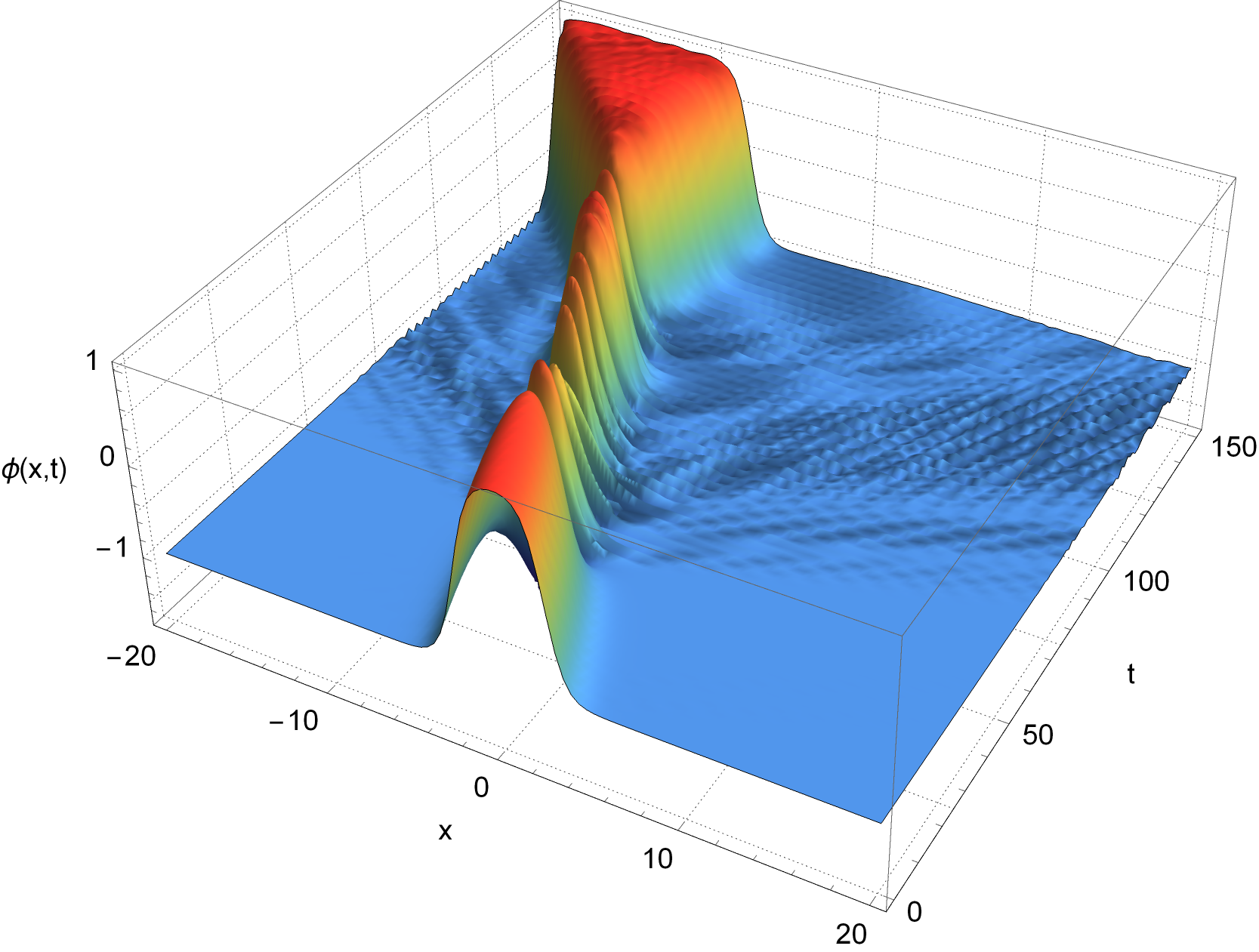}}\hfill
  \subfigure[The case $p=0.0$ and $v_{\text{in}}=0.196.$]{\includegraphics[height=4.33cm,width=5.33cm]{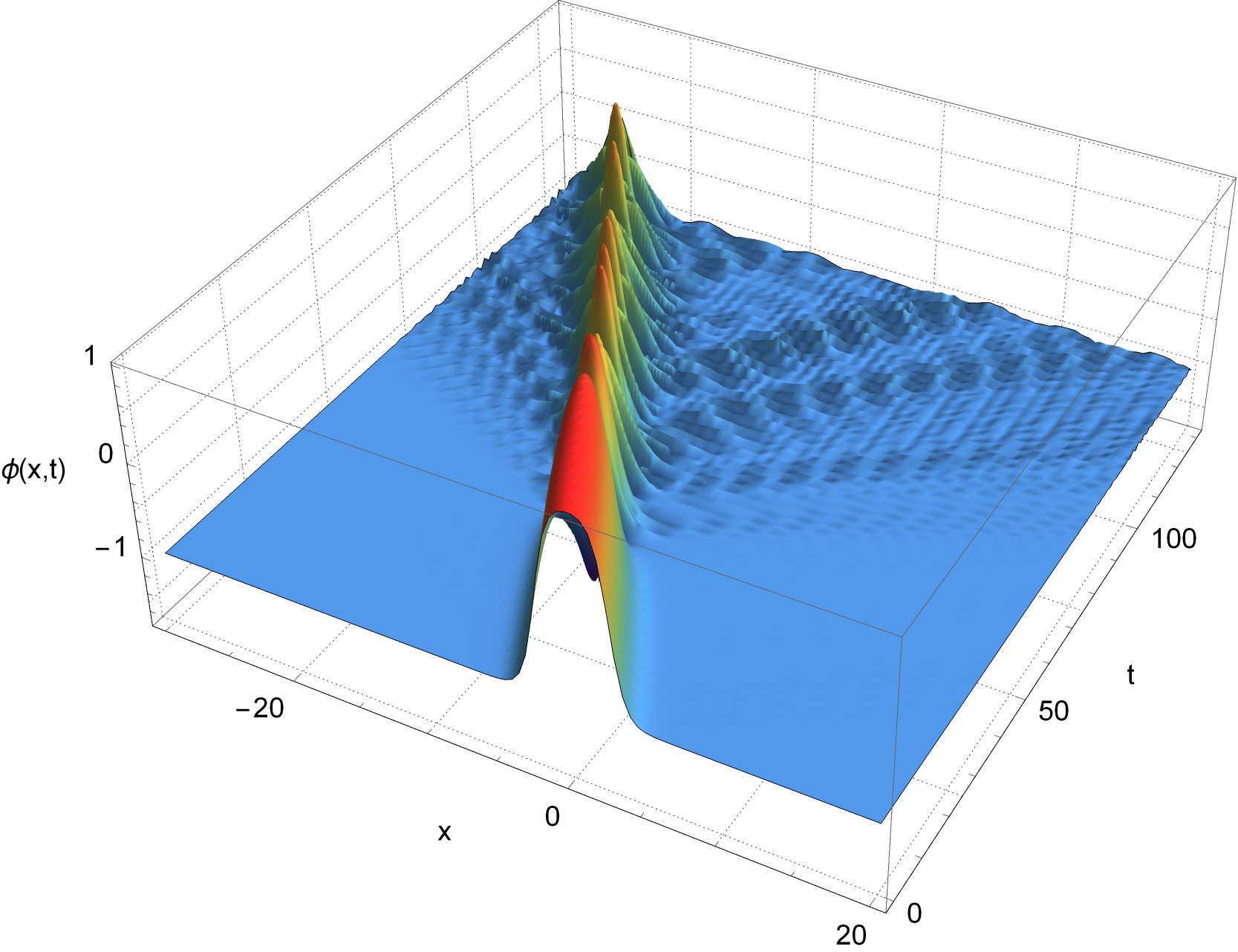}}\\
  \subfigure[The case $p=0.1$ and $v_{\text{in}}=0.049.$]{\includegraphics[height=4.43cm,width=5.33cm]{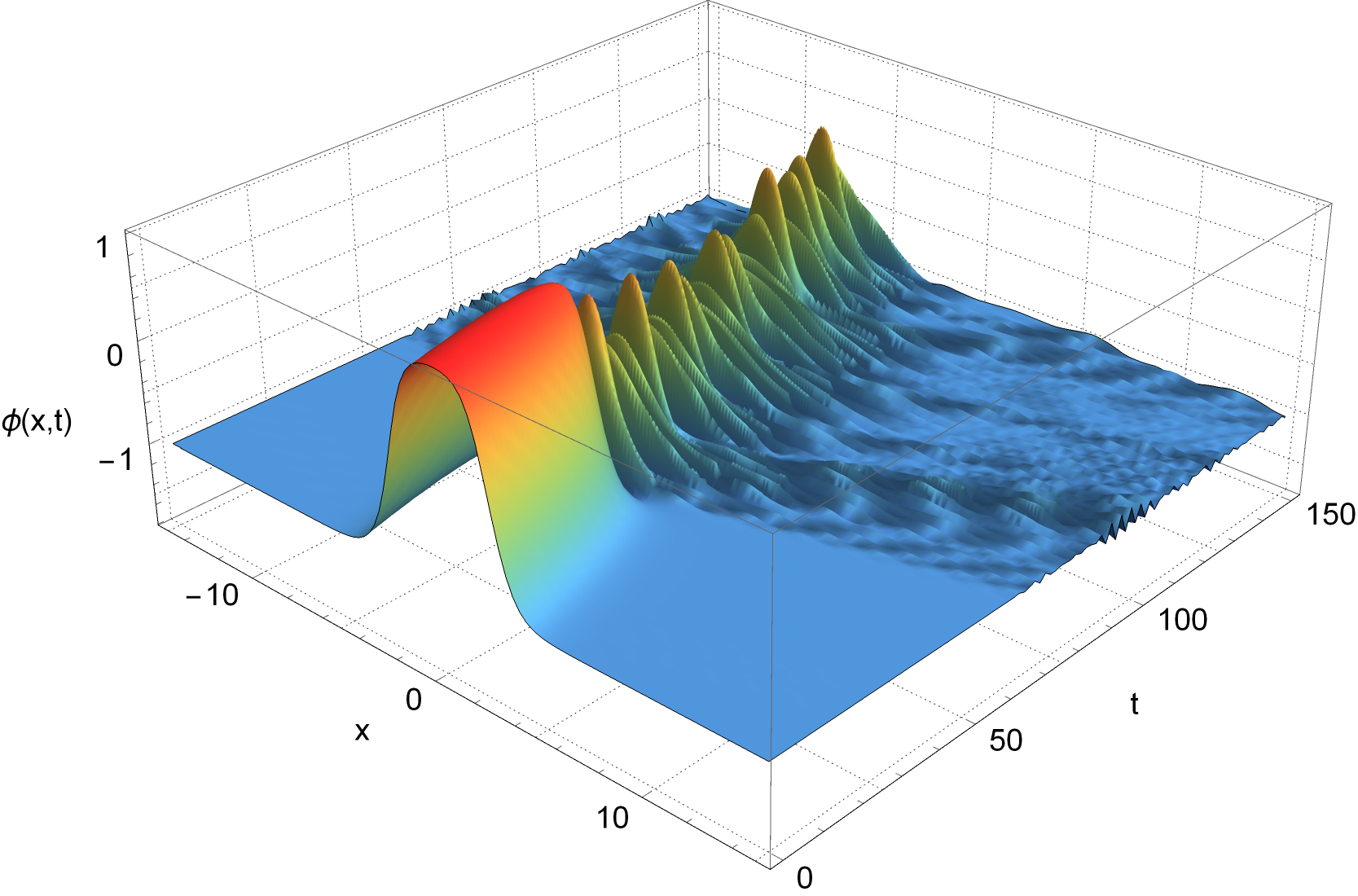}}\hfill
  \subfigure[The case $p=0.1$ and $v_{\text{in}}=0.098.$]{\includegraphics[height=4.43cm,width=5.33cm]{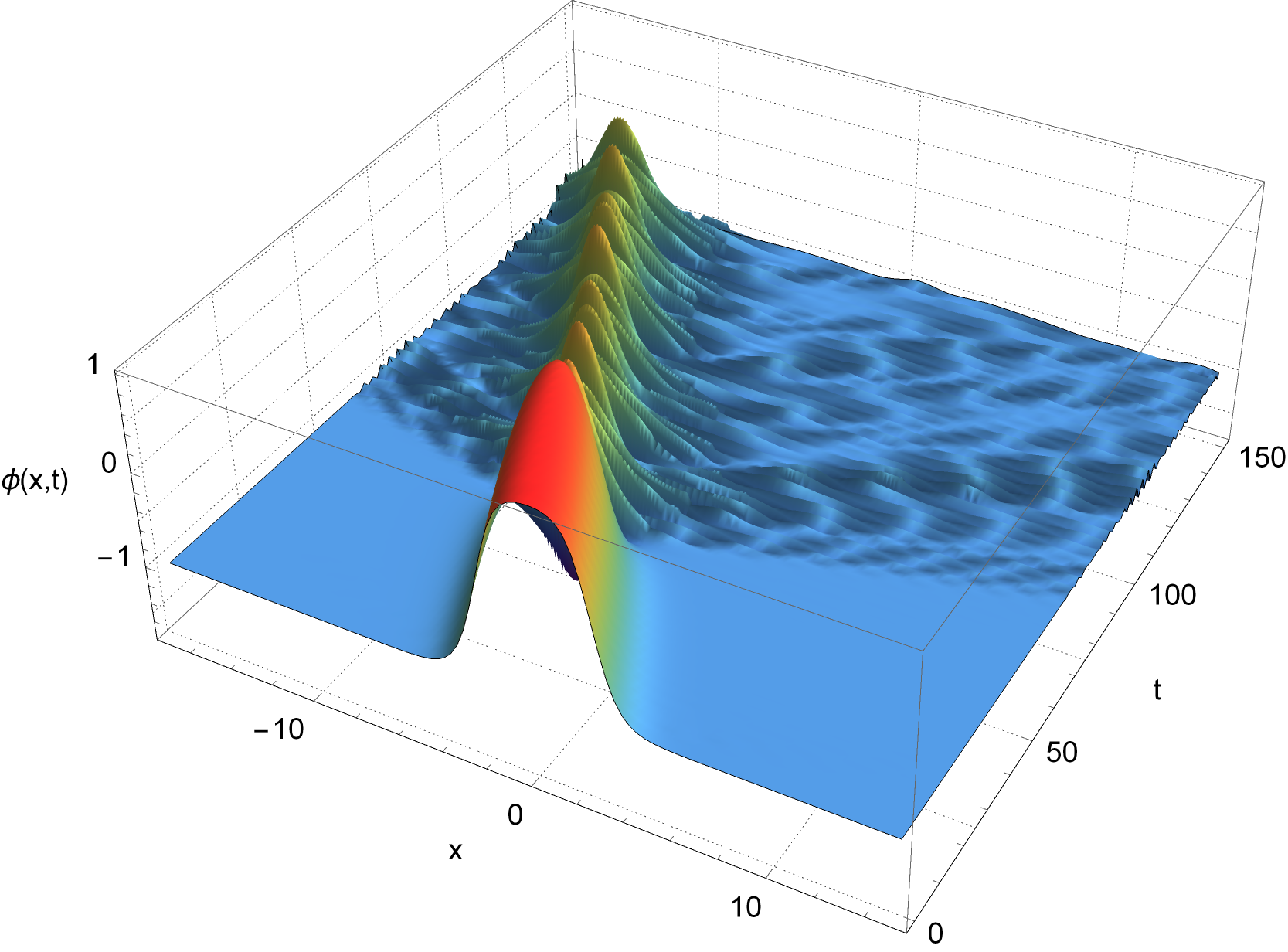}}\hfill
  \subfigure[The case $p=0.1$ and $v_{\text{in}}=0.196.$]{\includegraphics[height=4.33cm,width=5.33cm]{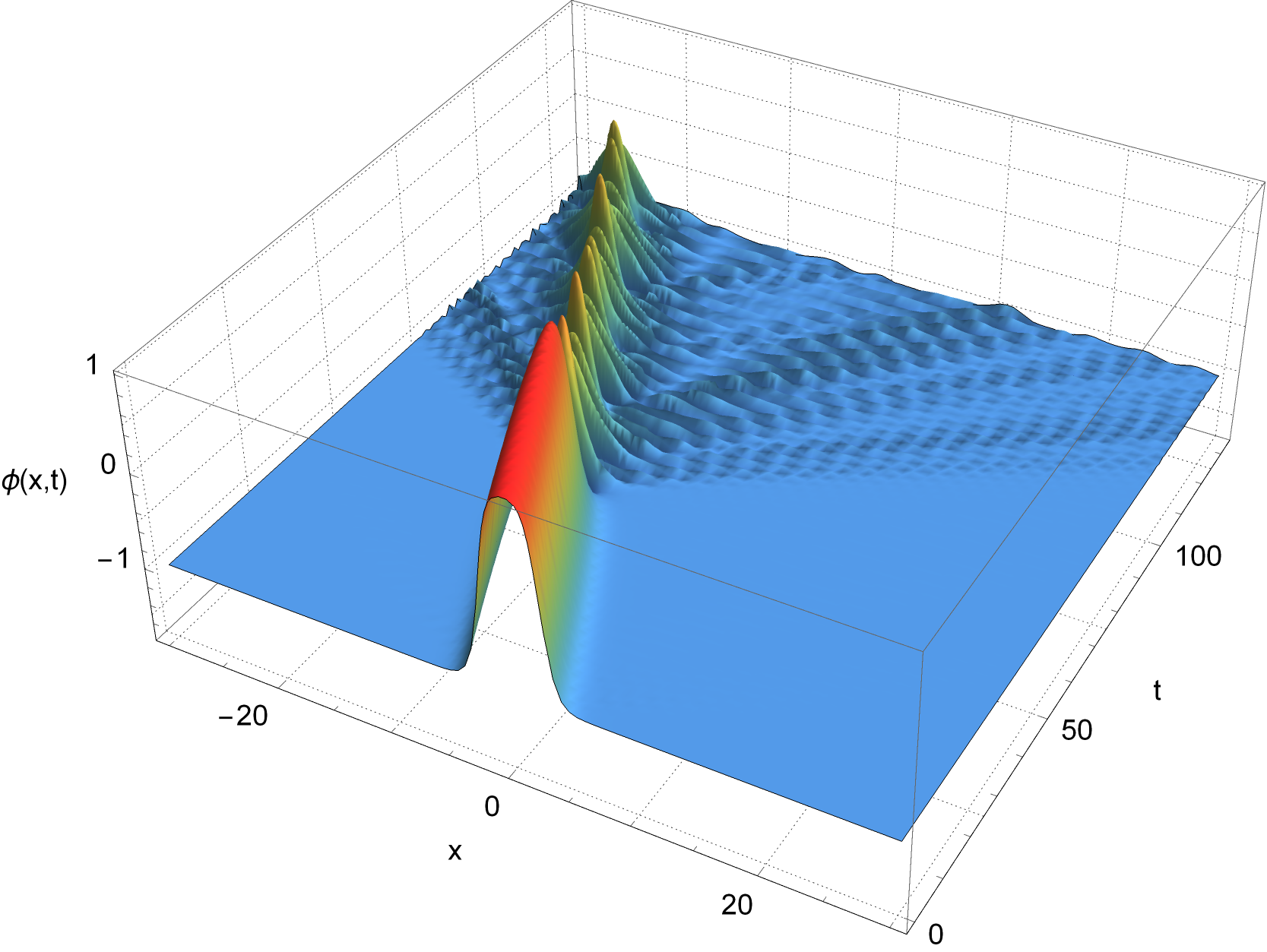}}\\
    \subfigure[The case $p=0.2$ and $v_{\text{in}}=0.049.$]{\includegraphics[height=4.3cm,width=5.33cm]{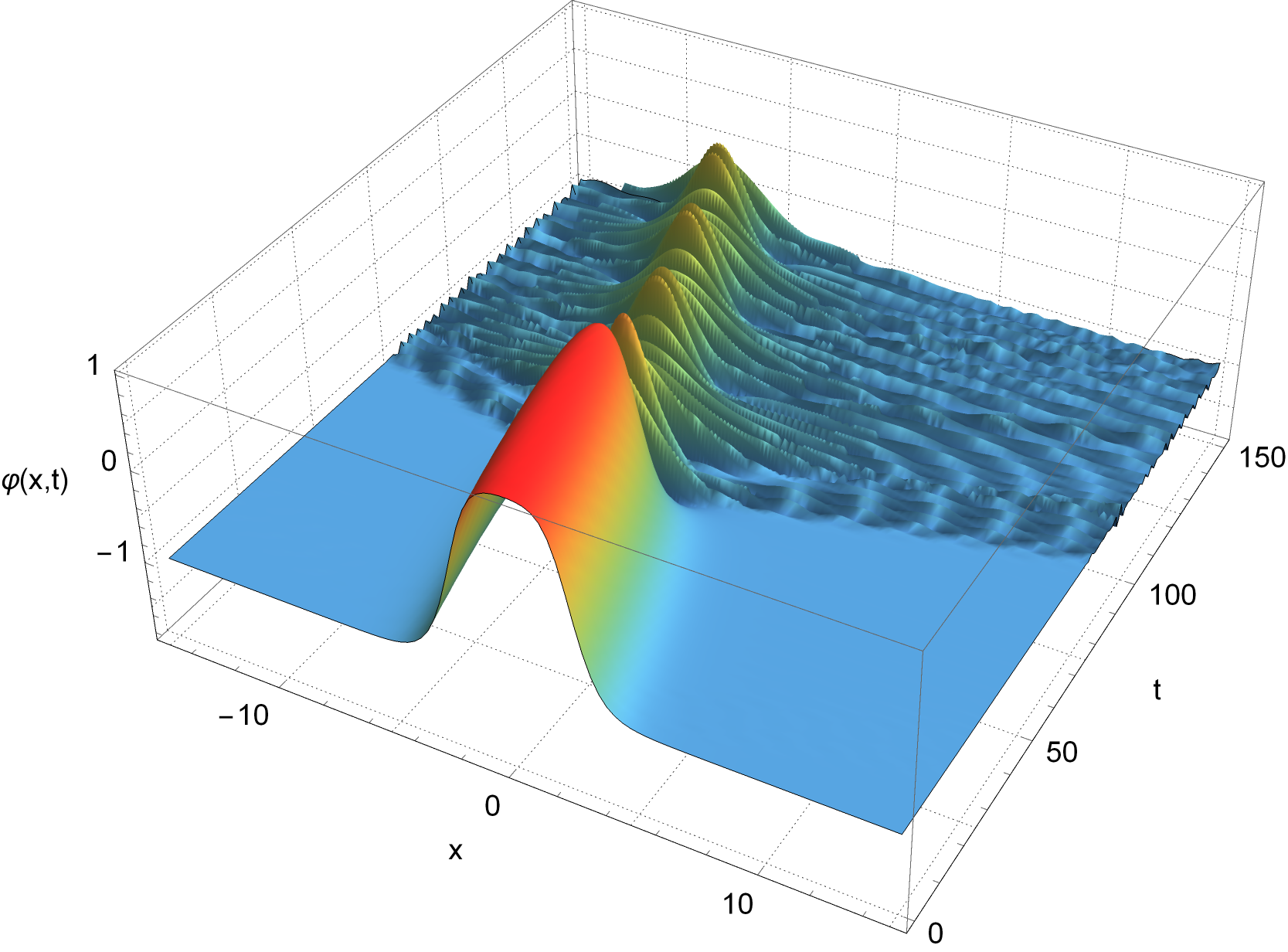}}\hfill
  \subfigure[The case $p=0.2$ and $v_{\text{in}}=0.098.$]{\includegraphics[height=4.33cm,width=5.33cm]{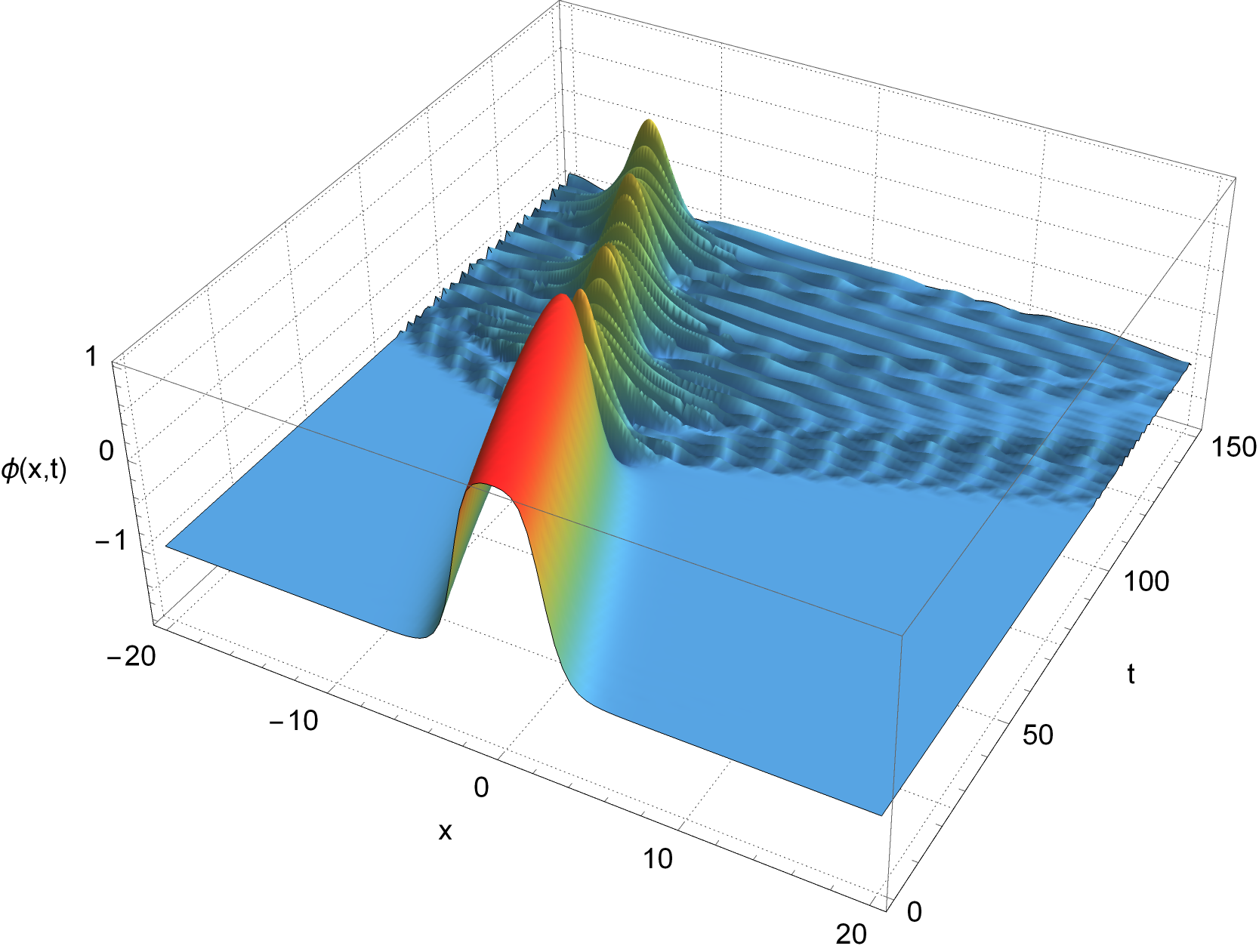}}\hfill
  \subfigure[The case $p=0.2$ and $v_{\text{in}}=0.196.$]{\includegraphics[height=4.33cm,width=5.33cm]{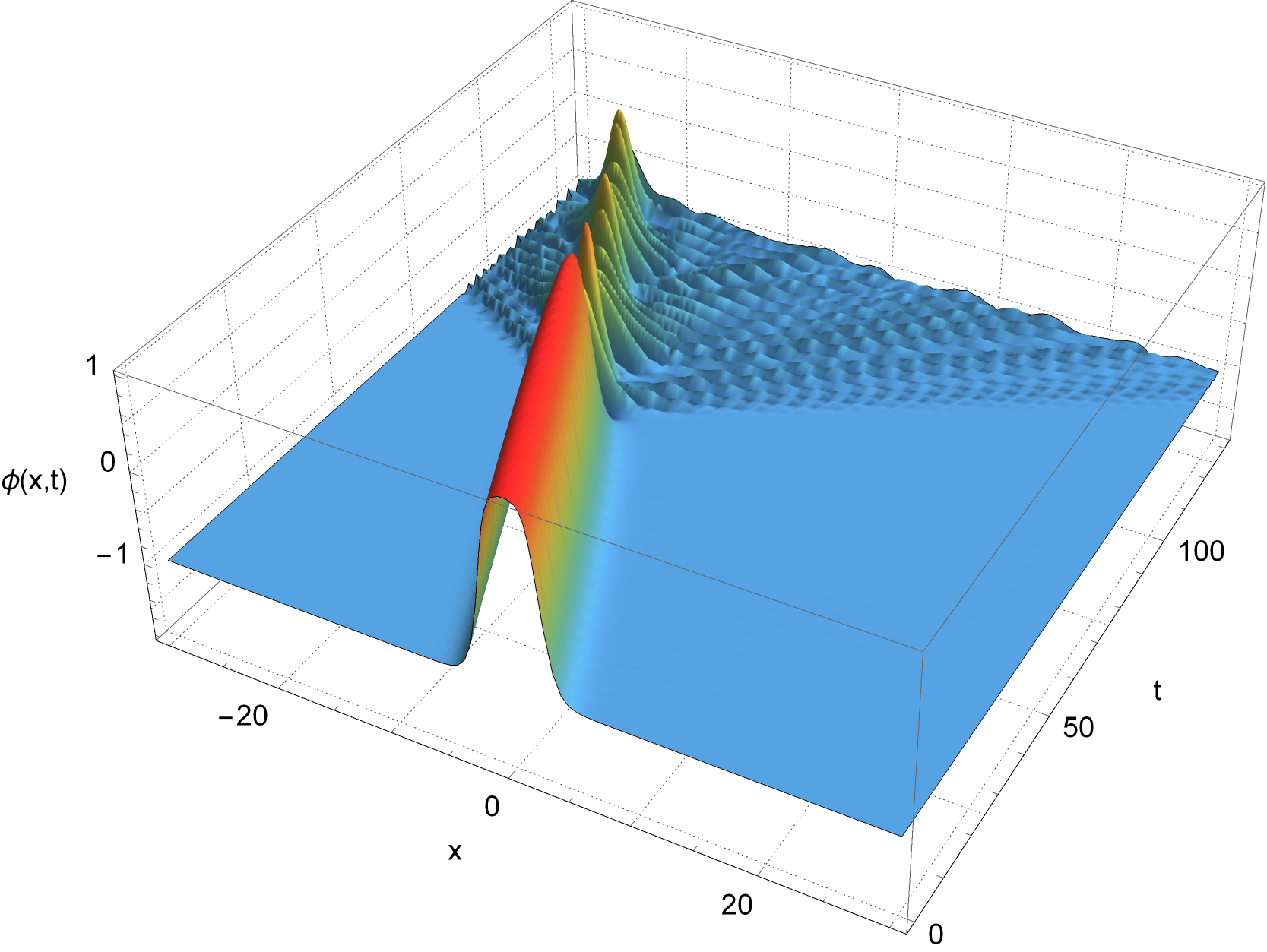}}
\caption{Formation of a bion in kink-antikink collisions at $v_{\text{in}}\leq v_{\text{cr}}$.}  \label{fig9}
\end{figure}

Furthermore, numerical inspection reveals that the configuration interpolating between $(-1,1,-1)$ significantly decreases amplitude when the critical velocity\footnote{In this framework, we refer to the critical velocity ($v_{\textrm{cr}}$) as the velocity at which the wave amplitudes reduce significantly.} satisfies $v_{\textrm{cr}} \geq 0.196$, except for the usual case, i.e., $p=0$. Indeed, due to the presence of asymmetry (with a compact-type profile in the range $0<\phi\leq\nu$), a dissipation process occurs, leading to the disappearance of kink/antikink configurations for specific $p$ values, e.g., $p>0$. In this conjecture, when scattering occurs, instead of simply reflecting or propagating after the interaction (as shown in Fig. \ref{fig9}(b)), the structures recombine, forming several secondary waves with smaller amplitudes. This phenomenon arises from the dissipation of particle energy as the solutions increase the asymmetry produced by the contribution $\textrm{e}^{2p\phi}$ ($p>0$) [see Figs. \ref{fig9}[(a)-(i)]].

We highlight that, within the range of initial velocities $v_{\textrm{in}}< v_{\textrm{cr}}$, no resonance phenomena, such as escape windows, were detected. Such phenomena could indicate resonant energy exchange between the translational and vibrational modes. For further details, see Refs. \cite{VAGani,Belendryasova}.

\subsection{The case $p< 0$}\label{secIVb}

Meanwhile, the evolution of the initial kink/antikink-like configurations $(-1,1,-1)$ is also obtained numerically by adopting the equation of motion \ref{Eq3} with asymmetry on the right. The asymmetry on the right induces a shift in that direction (a smaller shift) during the evolution of the structures, with the presence of bions becoming more pronounced.

\begin{figure}[!ht]
  \centering
  \subfigure[The case $p=-0.1$ and $v_{\text{in}}=0.049.$]{\includegraphics[height=4.33cm,width=5.33cm]{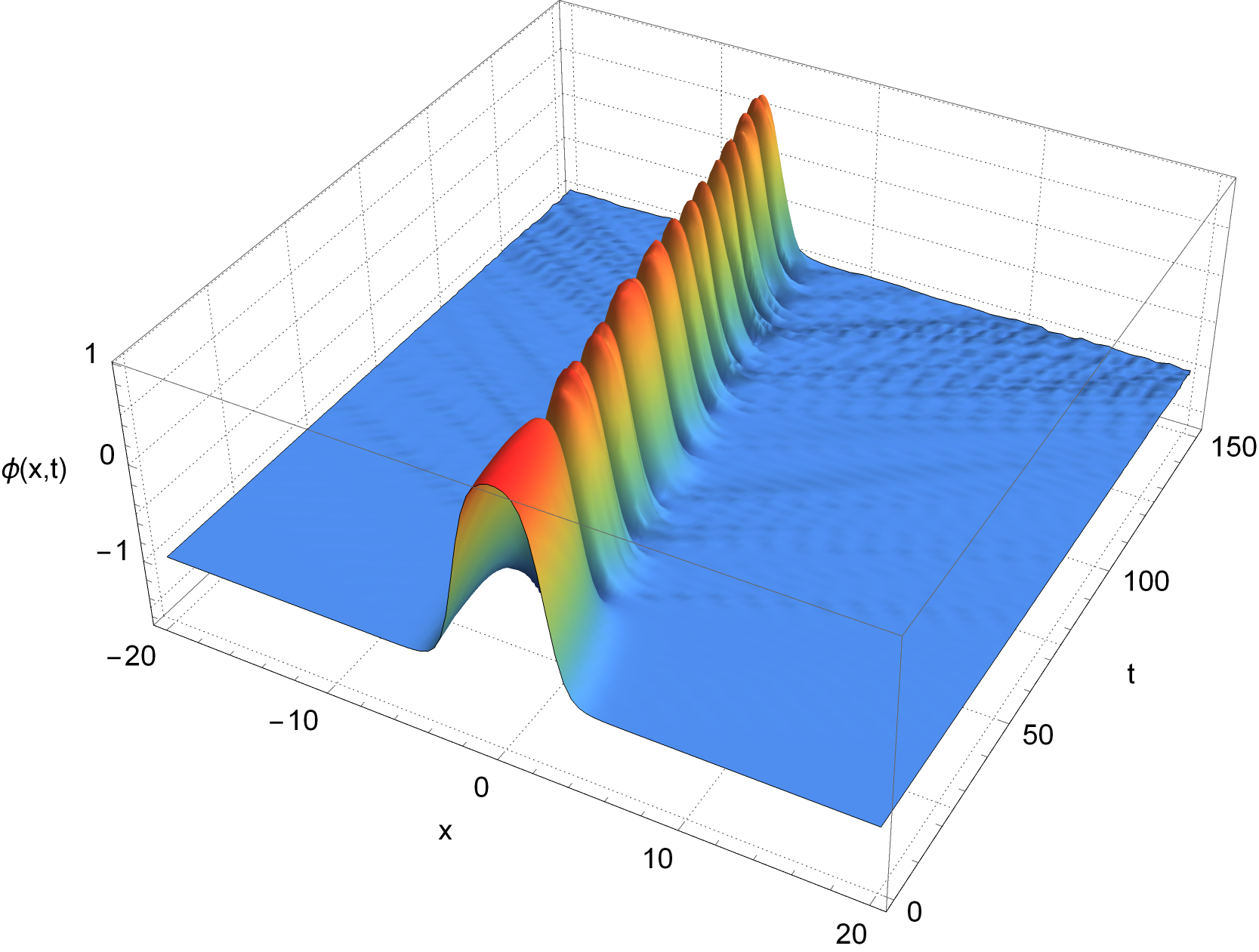}}\hfill
  \subfigure[The case $p=-0.1$ and $v_{\text{in}}=0.098.$]{\includegraphics[height=4.33cm,width=5.33cm]{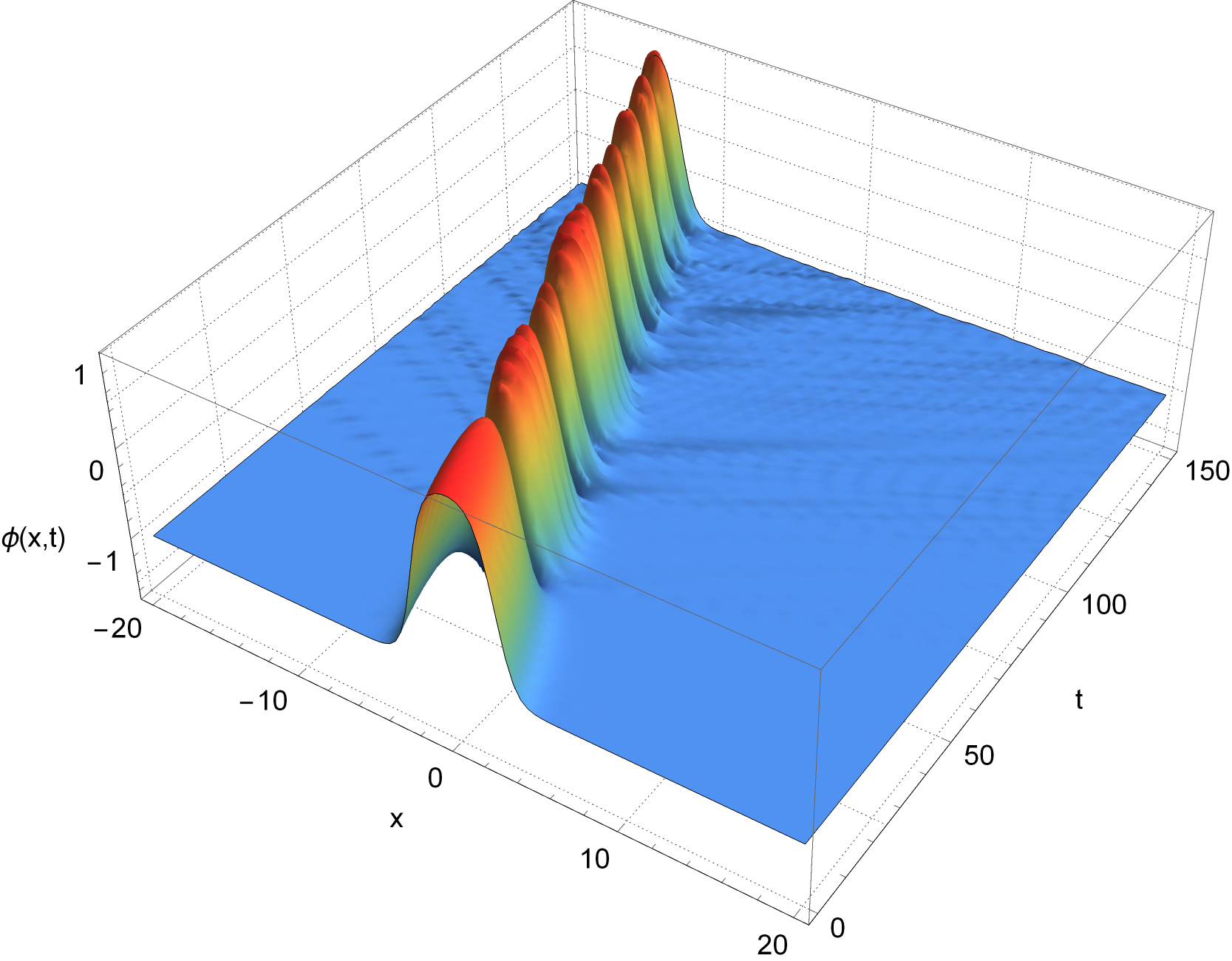}}\hfill
  \subfigure[The case $p=-0.1$ and $v_{\text{in}}=0.196.$]{\includegraphics[height=4.33cm,width=5.33cm]{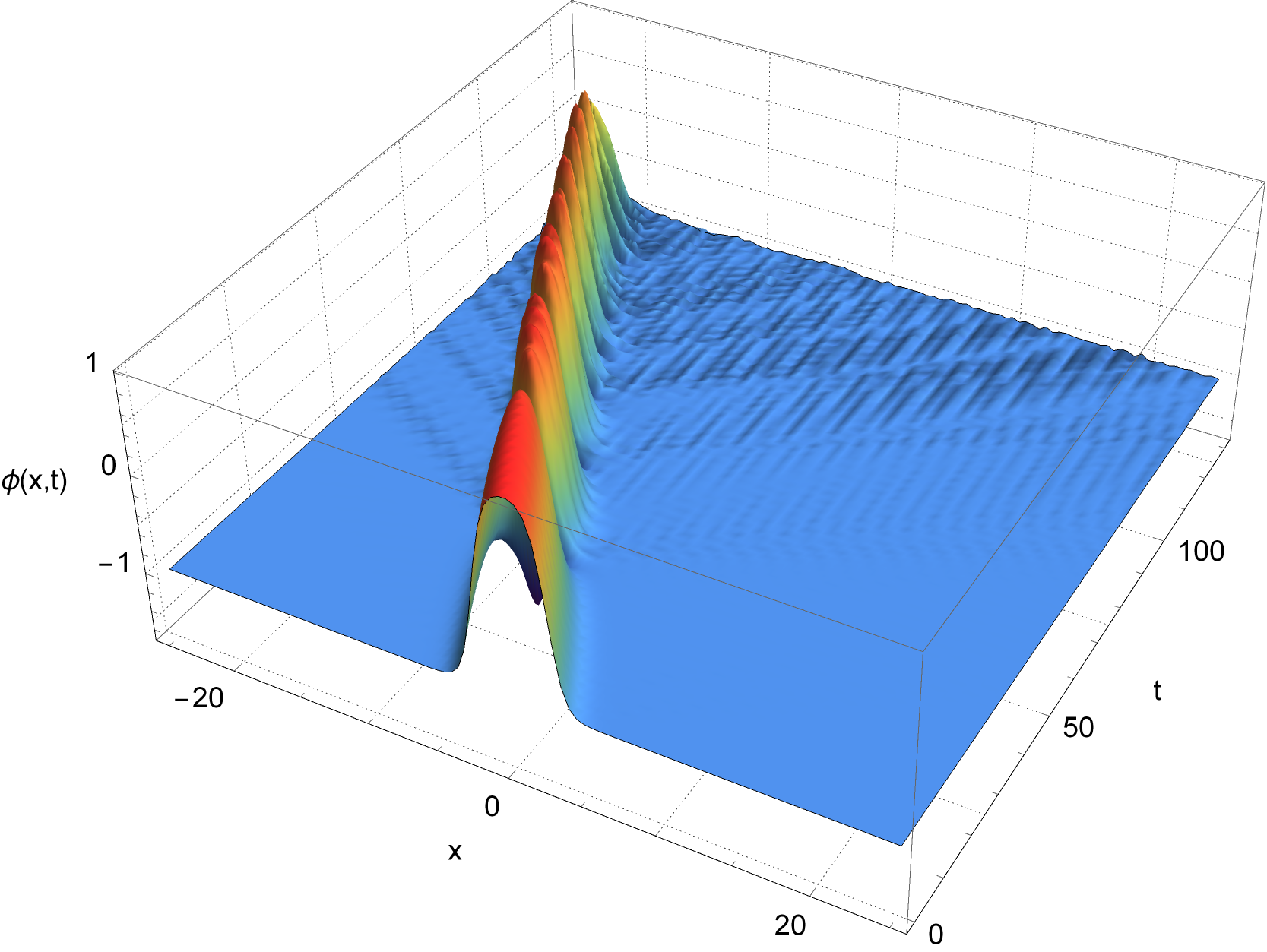}}\\
    \subfigure[The case $p=-0.2$ and $v_{\text{in}}=0.049.$]{\includegraphics[height=4.33cm,width=5.33cm]{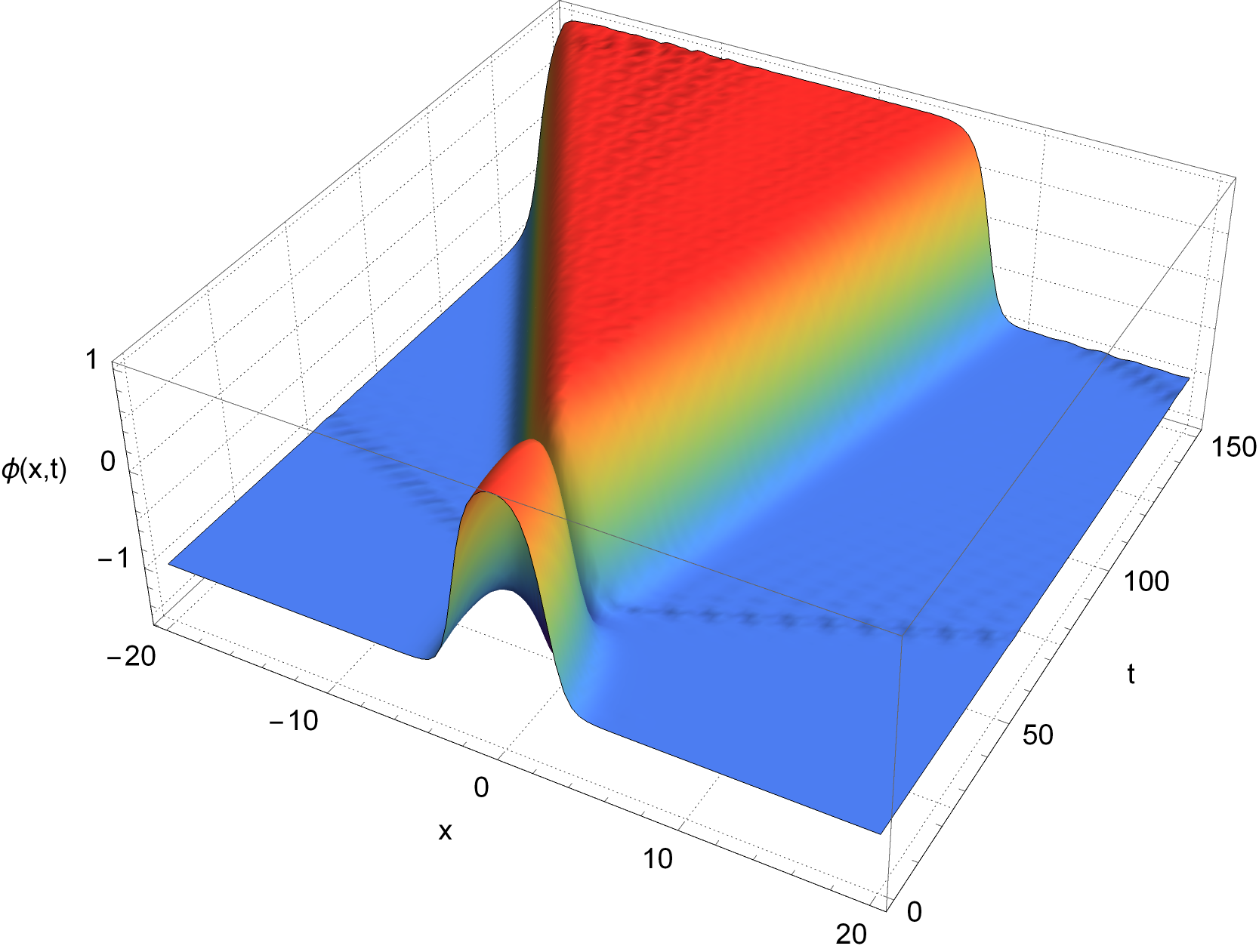}}\hfill
  \subfigure[The case $p=-0.2$ and $v_{\text{in}}=0.098.$]{\includegraphics[height=4.33cm,width=5.33cm]{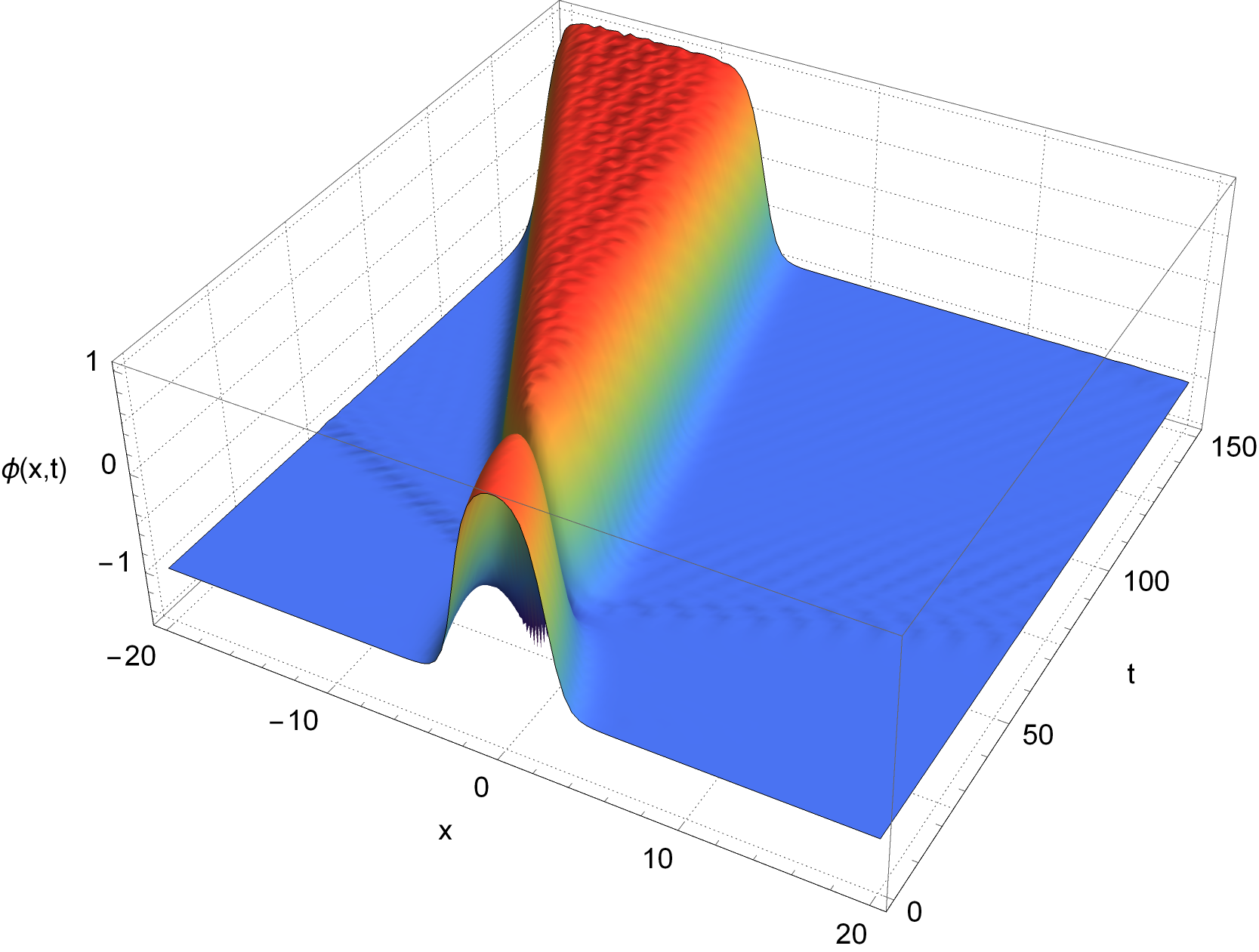}}\hfill
  \subfigure[The case $p=-0.2$ and $v_{\text{in}}=0.196.$]{\includegraphics[height=4.3cm,width=5.33cm]{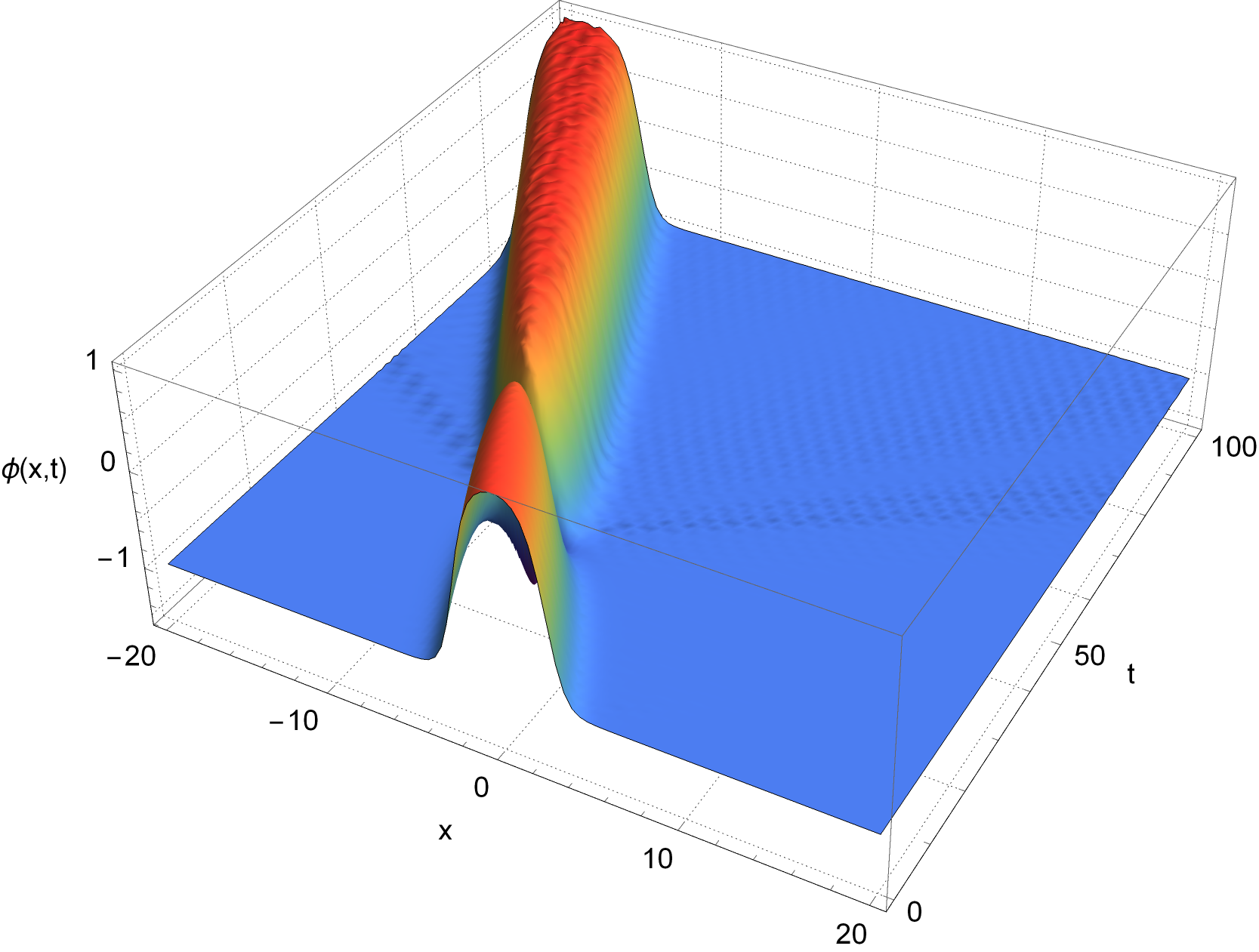}}
\caption{Formation of a bion in kink-antikink collisions at $v_{\text{in}}\leq v_{\text{cr}}$.}\label{fig10}
\end{figure}

Furthermore, numerical inspection reveals that the configuration also interpolating between $(-1,1,-1)$ annihilates when the critical velocity satisfies $v_{\textrm{cr}} \geq 0.196$ when $p=0.1$. Conversely, for $v_{\textrm{in}} < v_{\textrm{cr}}$, the kinks/antikinks annihilate, radiating their energy outward in the form of small-amplitude waves [see Figs. \ref{fig9}[(a)-(i)]]. One notes also that these small amplitude waves carry part of the kinetic energy. However, for $p\geq 0.2$ and $v_{\textrm{in}} > v_{\textrm{cr}}$, one notes inelastic reflections of kinks arise, see Figs. \ref{fig10}[(d)-(f)]. In this framework, we also observe that small amplitude waves carry away a part of the kinks’ kinetic energy.

\section{Summary and conclusion}

In this work, we adopted a Lorentz-invariant scalar theory with a modified $\phi^4$ potential by an exponential global impurity. This modification introduces a vacuum asymmetry in the theory. Naturally, one recovers the standard $\phi^4$ theory when the asymmetric contributions become negligible (i.e., $p\ll 1$). The asymmetry introduced in the potential suggests that, as $p\to 0$, the case with asymmetry to the right will have only trivial field configurations (i.e., $\phi=\phi_0\equiv$constant). In contrast, for the case with asymmetry on the left, topological solutions will always exist for any value of the parameter $p$ (i.e., the asymmetry parameter).

Naturally, when analyzing the solutions obtained for the studied cases, one notes that the configurations exhibit a shift at the critical point of energy density, moving away from the center of the kink/antikink-like solutions. This behavior suggests that the kink/antikink-like field configurations are asymmetric. Furthermore, we noted the existence of asymmetry in the profiles of the zero modes concerning field configurations, which ensures the translational invariance of the structures. In other words, the presence of the asymmetric zero modes ensures the shift of the kink/antikink-like configurations without incurring any energetic cost.

Finally, we observed that, when analyzing the system's dynamics, the formation of bions resulting from the collisions between kink and antikink involves an exchange of energy and momentum with the radiation emitted in the form of secondary waves (see Fig. \ref{fig9} and \ref{fig10}). This exchange appears to be directly related to the system's asymmetry, i.e., the variation of the parameter $p$.%Furthermore, one notes the existence of a violation in the conservation of the total momentum. One can interpret this violation of the total momentum as a direct consequence of the dissipative asymmetric term.

\section{Acknowledgment}

F. C. E. Lima expresses their gratitude to the Conselho Nacional de Desenvolvimento Cient\'{i}fico e Tecnol\'{o}gico (CNPq), grant 171048/2023-7 (CNPq/PDJ), for their invaluable financial support.

\end{document}